\documentclass[prl,twocolumn,superscriptaddress]{revtex4-1}
\usepackage{amsmath,amssymb,mathrsfs}
\usepackage{graphicx}

\begin{document}

\title{Exactly Solvable Points and Symmetry-Protected Topological Phases of Quantum Spins on a Zig-Zag Lattice}

\author{Haiyuan Zou}
\affiliation{Tsung-Dao Lee Institute, Shanghai Jiao Tong University, Shanghai 200240, China}

\author{Erhai Zhao}
\affiliation{Department of Physics and Astronomy and Quantum Materials Center,
 George Mason University, Fairfax, Virginia 22030, USA}

\author{Xi-Wen Guan}
\affiliation{State Key Laboratory of Magnetic Resonance and Atomic and Molecular Physics,
Wuhan Institute of Physics and Mathematics, Chinese Academy of Sciences, Wuhan 430071, China}

\author{W. Vincent Liu}
\affiliation{Department of Physics and Astronomy, University of Pittsburgh, Pittsburgh, Pennsylvania 15260, USA}
\affiliation{Wilczek Quantum Center, School of Physics and Astronomy and T.D. Lee Institute, Shanghai Jiao Tong University, Shanghai 200240, China}
\affiliation{Shenzhen Institute for Quantum Science and Engineering and Department of Physics, Southern University of Science and Technology, Shenzhen 518055, China}

\begin{abstract}
A large number of symmetry-protected topological (SPT) phases have been hypothesized for strongly interacting spin-1/2 systems in one dimension.
Realizing these SPT phases, however, often demands fine-tunings hard to reach experimentally. And the lack of 
analytical solutions hinders
the understanding of their many-body wave functions.
Here we show that two kinds of SPT phases naturally arise 
for ultracold polar molecules confined in a zigzag optical lattice. This system,  motivated by recent experiments,
is described by a spin model whose exchange couplings can be tuned by an external field to reach parameter regions not
studied before for spin chains or ladders.
Within the enlarged parameter space, we find the ground state wave function can be obtained exactly along a line and at a special point,
for these two phases respectively.
These exact solutions provide a clear physical picture for the SPT phases and
their edge excitations.
We further obtain
the phase diagram by using infinite time-evolving block decimation, and discuss the phase transitions between the two
SPT phases
and their experimental signatures.
\end{abstract}

\maketitle
The ground states
of strongly interacting many-body systems of quantum spins can differ from each other by three mechanisms: symmetry breaking, long range entanglement (topological order), or symmetry fractionalization~\cite{chen2011complete}. Symmetry-protected topological (SPT) phases are equivalent classes of states that share the same symmetries but are topologically distinct  \cite{chen2013symmetry,senthil2015symmetry,chen2011complete}. They only have short-range entanglement, are gapped in the bulk, but have edge or surface states protected by symmetries. Recent years have witnessed significant advancement in our understanding of fermionic and bosonics SPT phases. For example, for one-dimensional (1D) spin systems, a complete classification of possible SPT phases was achieved based on group cohomology~\cite{chen2011complete}. A plethora of SPT phases are shown to be mathematically allowed. When translational symmetry, inversion, time reversal (TR), and $D_2$ symmetry of spin rotation $\pi$ are all present, there are in total $2^{10}$ possible SPT phases in 1D \cite{chen2011complete}.

Only a small fraction of these SPT phases have been identified to arise from realistic spin models that are experimentally accessible.
The best known example is the Haldane phase of spin-1 antiferromagnetic Heisenberg chain~\cite{Haldane}. For spin-$1/2$ systems, spin ladders, $J_1-J_2$ chains with frustration (for example with antiferromagnetic next-nearest-neighbor interaction $J_2>0$) have been extensively studied~\cite{Furukawa2012,White1,alterchain,Kanter,FMleg}, but the parameter space explored was focused on solid state materials such as copper oxides
\cite{Furukawa2012}. For example, four SPT phases $D_\pm$, VCD$_\pm$ have been discussed in spin-$1/2$ chains~\cite{Ueda2014Chiral2}. And Ref.  \cite{liu2012symmetry} found four SPT phases $t_0, t_x, t_y, t_z$ in a spin-$1/2$ ladder and proposed ways to realize them using coupled quantum electrodynamics cavities. The $t_0$ and $t_z$ phases were also shown to exist in narrow regions for a ladder of dipole molecules~\cite{manmana2013topological}.
In quantum gas experiments, a noninteracting SPT phase was observed with fermionic ytterbium atoms~\cite{bSPT2018}, and
an interacting bosonic SPT phase was realized using Rydberg atoms~\cite{fSPTarxiv2018}.

In this Letter, we propose and solve a highly tunable 1D spin-$1/2$ zigzag lattice model describing  polar molecules~\cite{Yan:2013xe,PhysRevLett.113.195302,PhysRevLett.107.115301}  (or magnetic atoms \cite{PhysRevLett.111.185305}) localized in a deep optical lattice. This model has several appealing features as a platform to realize SPT phases. (1) It is inspired by recent experimental realization of spin-$1/2$ $XXZ$ model using polar molecules in optical lattices~\cite{Yan:2013xe,PhysRevLett.113.195302,PhysRevLett.107.115301}. (2) The relative magnitude and sign of the exchange interactions are relatively easy to control  by tilting the dipole moment using an electric field to reach {a large, unexplored parameter space}.
The frustration resulting from dipole tilting has been recently shown to give rise to possible spin liquid states in 2D~\cite{DSL2015,Our2017,keles2018absence,keles-prb}. (3)
The bulk of its phase diagram is occupied by two SPT phases, the singlet-dimer (SD) and even-parity dimer (ED) phase.  (4) The exact ground state wave function for each SPT phase is found and their nature is firmly established by exploiting the characteristic of the lattice as a chain of edge sharing triangles. 
The spin-1/2 edge states of an open chain are also derived. (5) It reveals a novel direct phase transition between the SPT phases. 

\textit{The model.---}
Our model, illustrated in Fig. 1, is a spin-$1/2$ $XXZ$ model on the one-dimensional zigzag chain,
\begin{equation}
H=\sum_{i,j}J_{i,j}[S^x_iS^x_j+S^y_iS^y_j+\eta S^z_iS^z_j].
\label{eq:model}
\end{equation}
Here $i,j$ are the site indices, and $\eta$ is the exchange anisotropy.
The exchange coupling is restricted to nearest neighbors (NN) and next nearest neighbors (NNN),
\begin{equation}
J_{2i,2i+1}=J_1,\hspace{0.5cm} J_{2i-1,2i}=J'_1,\hspace{0.5cm} J_{i,i+2}=J_2.
\end{equation}
So the NN exchange alternates between $J_1$ and $J_1'$ (see Fig. 1). In the special case of $\eta=1$, the model reduces to the $J_1$-$J_2$-type Heisenberg model with bond alternation ($J_1\neq J_1'$).
In the literature, the $XXZ$ chain or $J_1$-$J_2$ Heisenberg chain have been extensively studied~\cite{PhysRevB.63.174430,Furukawa2012}. It is known that when $J_2>0$ (assuming $\eta>0$), the system is frustrated. The model has a rich phase diagram on the plane spanned by the two independent parameters: $\eta$ and $J_1/J_2$~\cite{Furukawa2012}. With 
a small bond alternation $\delta=|J_1-J_1'|\ll |J_1|,|J_1'|$, there are four SPT phases~\cite{Ueda2014Chiral,Ueda2014Chiral2}. The parameter space of this model,
e.g. for $J_2<0$ and strong bond alteration $\delta\sim |J_1|,|J_1'|$, have not been explored~\footnote{In Ref. \cite{Furukawa2012}, a schematic phase diagram (Fig. 11) was conjectured based on bosonization.}.

The model Eq.~\eqref{eq:model} naturally arises for polar molecules such as KRb and NaK localized in deep optical lattices ~\cite{Yan:2013xe,PhysRevLett.113.195302,PhysRevLett.107.115301}. Here the spin 1/2 refers to two chosen rotational states of the molecules, and the exchange interaction $J_{i,j}$ is dictated by the dipolar interaction between the two dipoles, which depends on their relative position $\mathbf{r}_{ij}=\mathbf{r}_i-\mathbf{r}_j$
as well as $\hat{d}$, the direction of the dipoles controlled by external electric field ~\cite{DSL2015,Our2017}. Explicitly, $J_{i,j}=J[1-3(\hat{r}_{ij}\cdot\hat{d})^2]/r_{ij}^3$ where $J>0$ sets the overall exchange scale.
For the zigzag lattice, we assume the external field is in plane, and makes an angle $\theta$ with the $y$ axis (Fig. 1). We further assume the lattice spacing is large and neglect longer range interaction beyond NNN. It follows that
\begin{eqnarray}
J_1&=&J[1-3\cos^2(\theta+\gamma)],\nonumber\\
J'_1&=&J[1-3\cos^2(\theta-\gamma)],\\
J_2&=&J[1-3\sin^2\theta]/8\sin^3\gamma. \nonumber
\end{eqnarray}
In general, the zig-zag angle $\gamma$ can be tuned.
Here we keep $\gamma=30^\circ$ fixed, so the zigzag lattice consists of a chain of identical, equilateral triangles.
Note that itinerant dipoles~\cite{Yelin2017} and atoms~\cite{Vekua2013,Jo2015} on the zig-zag lattice have been studied. Here we focus on {spin models of localized dipoles}.
The anisotropy $\eta$ can be tuned by varying the strength of the electric field \cite{Yan:2013xe,DSL2015}.

\begin{figure}[h]
\centering
\includegraphics[width=0.4\textwidth]{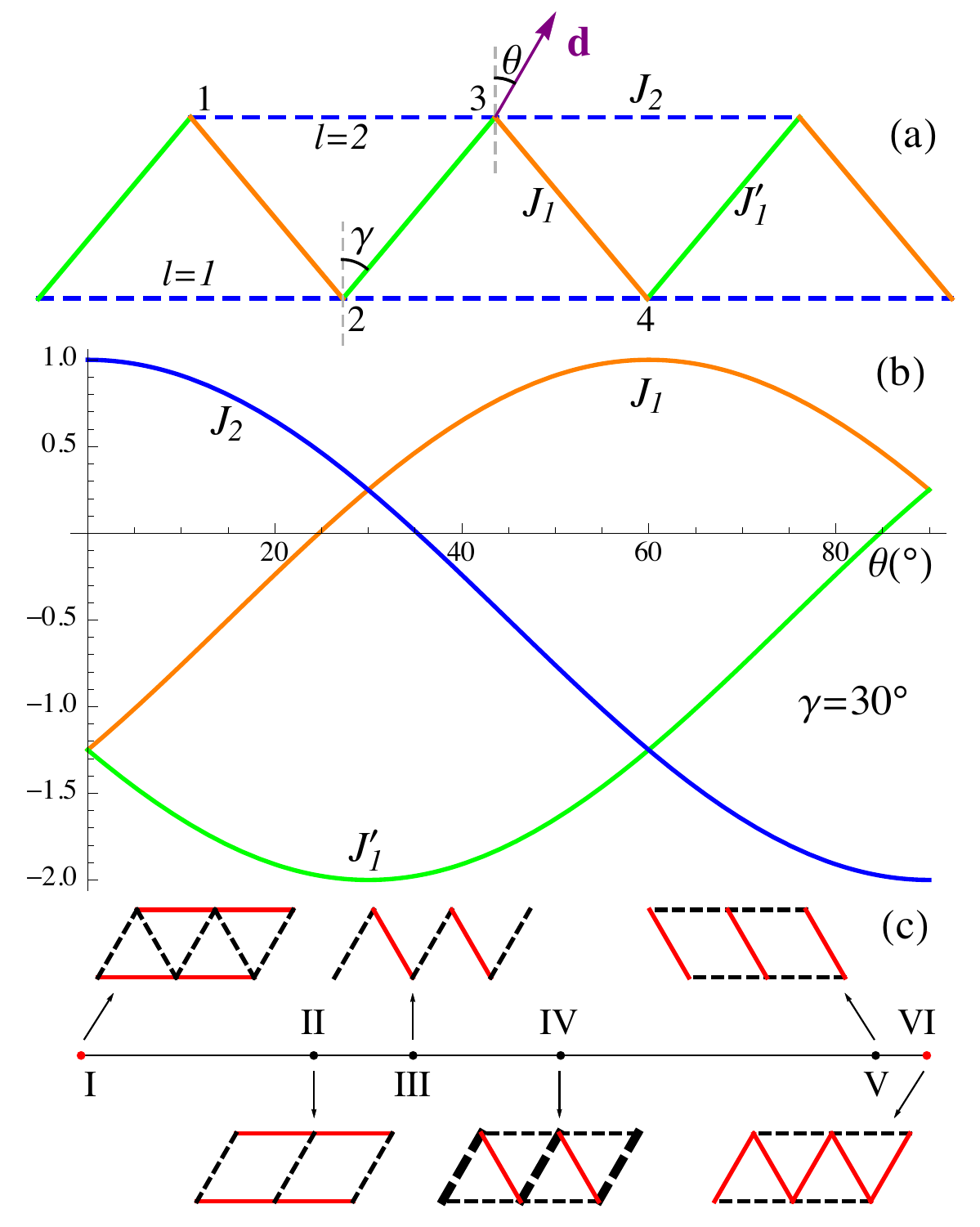}
\caption{(a) Dipolar molecules localized on a zigzag chain. The dipoles point to the $\mathbf{d}$ direction controlled by external electric field, forming an angle $\theta$ with the vertical direction.
The exchange couplings $J_1$, $J'_1$ and $J_2$ are defined in Eq. (2),
(b) The variation of the exchange couplings as functions of $\theta$ ($\gamma=30^\circ$).
(c) This highly tunable model contains a few limits, some of which studied before in the literature.
(I) a $J_1-J_2$ chain ~\cite{Furukawa2012}; (II) a coupled antiferromagnetic ladder~\cite{White1}; (III) a bond alternating chain~\cite{alterchain}; (V) a ferromagnetic ladder~\cite{FMleg}; (VI) weakly coupled ferromagnetic chains. The solid line and dashed line stand for positive (antiferromagnetic) and negative (ferromagnetic) coupling respectively.
The point (IV), where $J_1'=2J_2<0$ and $J_1>0$, is exactly solvable. Here the thick dashed line indicates $|J_1'|>|J_2|$.
 }
\label{fig:model}
\end{figure}

\textit{Tuning the exchange couplings.---}
By titling electric field (and the dipole moment $\hat{d}$), one sweeps through the parameter space of ${H}$ and gain access to nontrivial SPT phases. It is sufficient to consider $\theta\in[0,90^\circ]$. The resulting exchange coupling $J_1,J_1',J_2$ are shown in Fig.~\ref{fig:model}(b). As $\theta$ is varied, the system goes through a few points studied before in the literature. For example, at $\theta=0^\circ$, $J_1=J_1'=-1.25J_2$, the ground state was shown to be the so-called Haldane dimer phase~\cite{Furukawa2012}. At $\theta\sim 25^\circ$, $J_1=0$, the zigzag chain reduces to a ladder of ferromagnetically coupled antiferromagnetic chains, known to be connected to the spin-1 Haldane chain~\cite{White1}. At $\theta\sim 35^\circ$, where $J_2=0$, the system turns into a spin chain with alternating ferro- and antiferroexchange~\cite{alterchain}. At $\theta\sim 85^\circ$, $J'_1=0$, it reduces to a ladder system of two ferromagnetic chains with antiferromagnetic coupling and a ground state called the rung singlet phase~\cite{FMleg}. These ground states seem unrelated: they bear distinct names and are obtained using different methods for various models.

A main result of our work is that all the aforementioned points [Fig. 1(c)] belong to a single phase that extends to all $\theta\in [0^\circ,90^\circ)$ and $\eta=1$, and
are adiabatically connected to each other before touching the Tomonaga-Luttinger liquid (TLL) limit at $\theta=90^\circ$. Our model $H$ thus unifies these known topological phases in one-dimensional spin-1/2 systems.
Furthermore, we will show that the ground state wave function can be obtained exactly for a special point [IV in Fig. 1(c)] at $\theta\sim 50.9^\circ$, where $J_1'=2J_2<0$ and $J_1>0$. We prove that it is a pure product state of singlet dimers. Via continuity, the ground state of our model for $\eta=1$, including its topological character, can then be understood from this exact ground state. We will also show that as for $\eta<1$, a different SPT phase arises, and it also has an exactly solvable point.

\begin{figure}[h]
\centering
\includegraphics[width=0.45\textwidth]{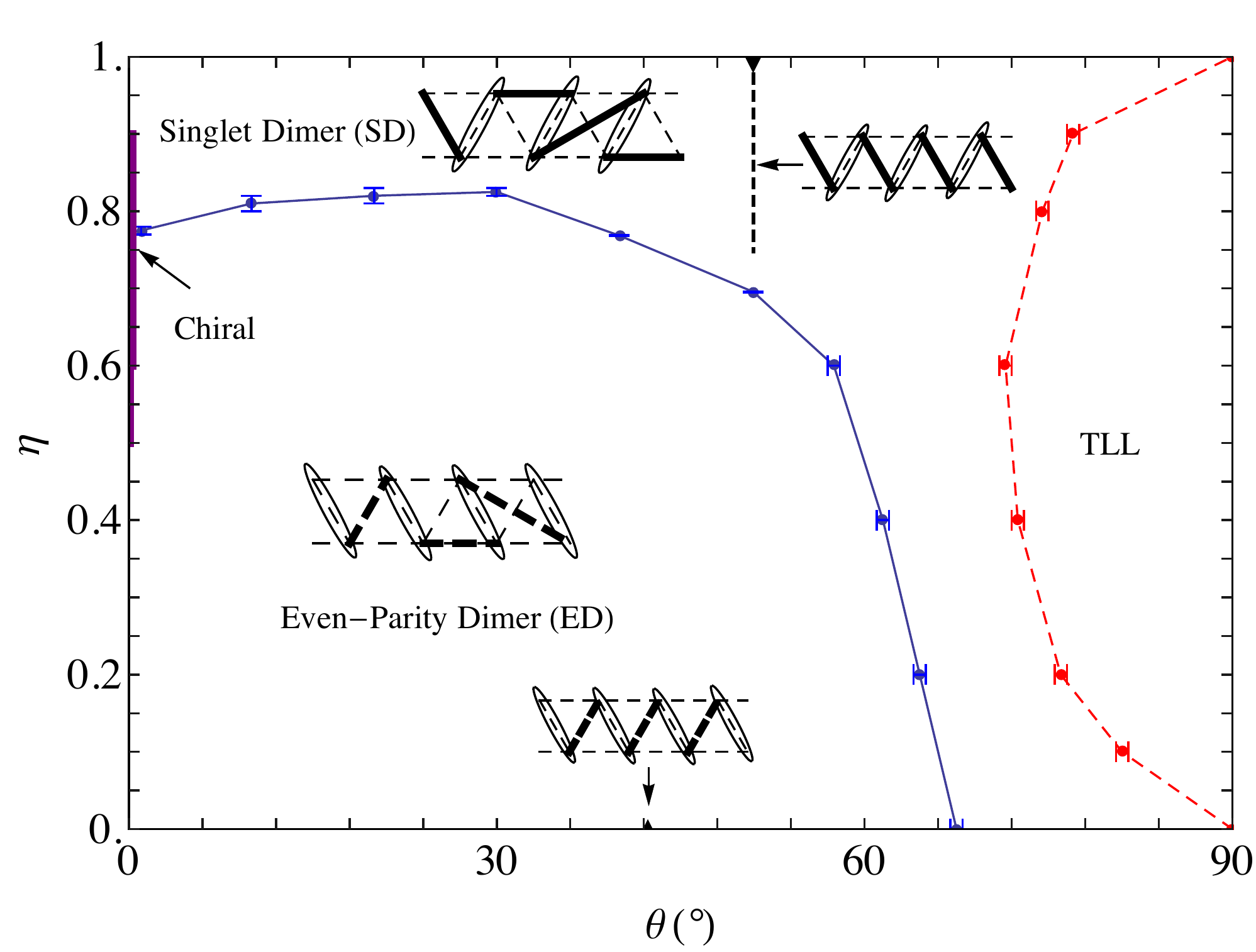}
\caption{The phase diagram of $H$ in the $\theta-\eta$ plane obtained from  iTEBD. The insets depict the singlet dimer (SD) and even-parity dimer (ED) phases. The thick solid lines in the SD case indicate a singlet $(\vert\uparrow\downarrow\rangle-\vert\downarrow\uparrow\rangle)/\sqrt{2}$ and the thick dashed lines in the ED case stand for an even-parity bond $(\vert\uparrow\downarrow\rangle+\vert\downarrow\uparrow\rangle)/\sqrt{2}$. The oval stands for the effective spin-1's defined in Eq.~\ref{eq:string}. The dashed line at $\theta\sim 50.9^\circ,\eta\in [0.747,1]$ and a point at $\eta=0,\theta\sim 42.4^\circ$ are exactly solvable, and their singlet/even-parity product state are shown. A chiral phase (Ch) exists in a small $\theta$ regime, and the Tomonaga-Luttinger liquid (TLL) phase occupies the large $\theta$ region. The error bars are due to the finite step size in scanning $\theta$ or $\eta$.
}
\label{fig:phaseD}
\end{figure}

{\it Phase diagram.---}To orientate the discussion, first we summarize the phase diagram of $H$ on the $\theta$-$\eta$ plane in Fig.~\ref{fig:phaseD}, obtained from infinite time-evolving block decimation (iTEBD) numerical calculations~\cite{iTEBD}. Here both the SD and ED phase
are gapped SPT phases, while the TLL phase is gapless. For a very narrow region, $\theta<0.5^\circ$, there is also a gapless chiral phase consistent with previous study~\cite{Ueda2014Chiral}. The chiral phase is not our main focus here and discussed further in the Supplementary Material ~\cite{noteS}. The suppression of the chiral phase is due to the alternating NN coupling which breaks the translational symmetry $S_i\rightarrow S_{i+1}$. In large $\theta$ region, the arc-shaped phase boundary between the SD and TLL phase on the $\theta$-$\eta$ plane is consistent with the prediction from effective field theory~\cite{noteS}.

The iTEBD method is based on the matrix product state representation of many-body wave functions in the thermodynamic limit. The Schmidt rank $\chi$ characterizes the entanglement of the system and it serves as the only adjustable parameter for precision control. Our calculation employs a unit cell of four sites and random complex initial states. Several quantities are computed to characterize the phases and detect possible phase transitions. The first is the string order parameter~\cite{string1989} defined as
\begin{equation}
{O}^z_n  =  - \! \lim_{r\to\infty}\langle(\hat{S}^z_{n} \! + \! \hat{S}^z_{n+1}) e^{ i \pi \! \sum_k \! \hat{S}^z_k }
(\hat{S}^z_{2r+n} \! + \! \hat{S}^z_{2r+n+1})\rangle,
\label{eq:string}
\end{equation}
where the $k$ sum is restricted to ${n+2\leq k \leq 2r+n-1}$.
The motivation behind this definition is that the two neighboring spins $\hat{S}^z_n+\hat{S}^z_{n+1}$ may form an effective spin-1 degree of freedom (represented by an oval in Fig. 2).
A finite ${O}^z$ detects hidden long range order. The SD (ED) phase is associated with a finite $O^z_n$ value for even (odd) site, say $n=2$ ($n=1$).
A clear ED-to-SD phase transition is observed in Fig. 3(b) as $\eta$ is varied.

\begin{figure}[h]
\centering
\includegraphics[width=0.45\textwidth]{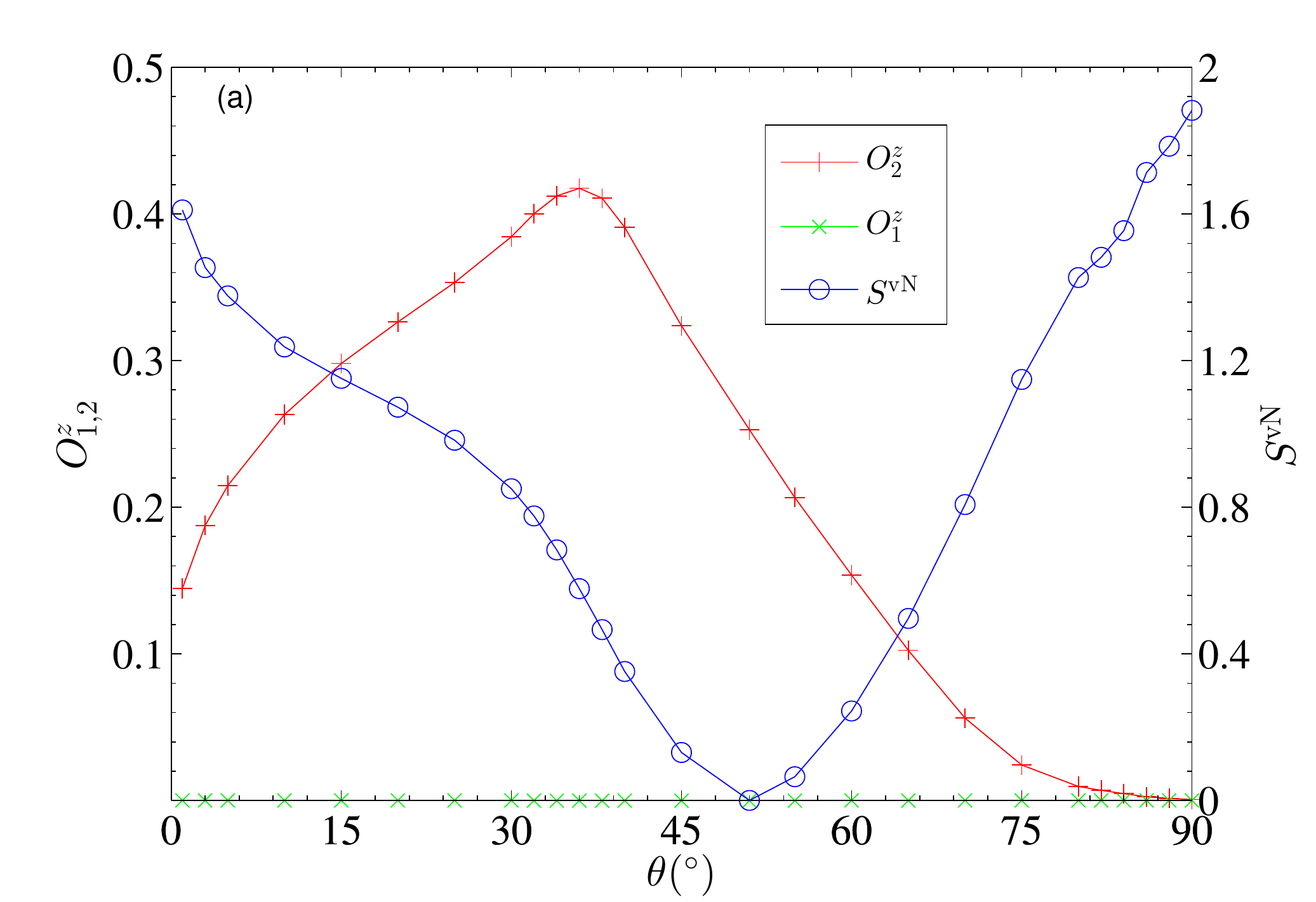}
\includegraphics[width=0.45\textwidth]{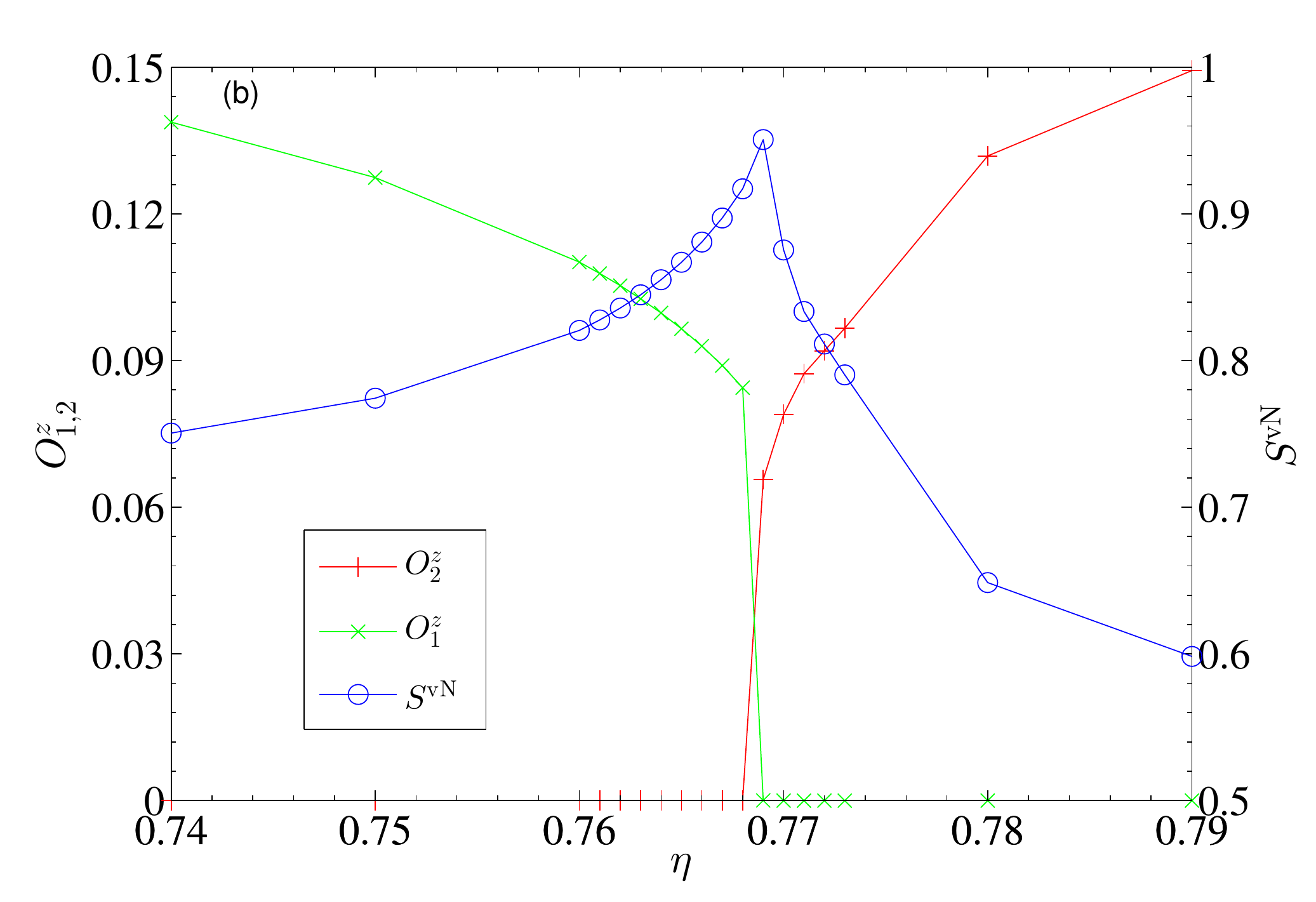}
\caption{Entanglement entropy $S^{\rm vN}$, string order parameter $O^z_1$ and $O^z_2$ for (a) the Heisenberg limit $\eta=1$, with $\chi=300$ and (b) along a line at $\theta=40^\circ$, with $\chi=100$. In (b), there is a phase transition at $\eta\sim0.769$ characterized by the peak of $S^{\rm vN}$ and jumps of string order parameters. }
\label{fig:Hei40}
\end{figure}

We also compute the von Neumann entanglement entropy by cutting a $J_1'$ bond,
$S^{\rm vN}=-\sum_{\chi} \lambda^2_{\chi}\ln\lambda^2_{\chi}$, where ${\lambda}_\chi$ is a set of normalized Schmidt coefficients with Schmidt rank $\chi$.
As shown in Fig.~\ref{fig:Hei40}(a) for $\eta=1$, both $S^{\rm vN}$ and $O^z_2$ are finite and continuous while $O^z_1$ remains zero as $\theta$ is tuned.
Together with other physical quantities~\cite{noteS},
these results show that the ground state remains in a single SD phase for all $\theta< 90^\circ$.
Interestingly, at $\theta=50.9^\circ$, $S^{\rm vN}$ vanishes, hinting a pure product state. We will show below that this is an exact solvable point.
On the other hand, as $\eta$ is varied for fixed $\theta=40^\circ$, $S^{\rm vN}$ develops a sharp peak in Fig.~\ref{fig:Hei40}(b). The peak position coincides with the jump in $O^z_n$ and
unambiguously identifies a phase transition from ED to SD. The variation of the string order parameters near the transition depends on the value of $\theta$. For large $\theta$, the transition appears to be first order, but it slowly changes to a continuous transition as $\theta$ decreases. We find that the central charge $c\sim 2$ at $\theta=10^\circ,20^\circ,30^\circ$, which suggests that the SPT phase transition 
has stronger interacting behavior than the Gaussian-type phase transition~\cite{noteS}.

\textit{Exact solutions.---}
Now we elucidate the nature of the SD and ED phases by two types of solvable points on the $\eta-\theta$ plane. At $\theta\sim 50.9^\circ$, the relation $J_1'=2J_2<0$ is satisfied with $J_1>0$.
Along this line (vertical dashed line in Fig. 2) of fixed $\theta$, the ground state of $H$ can be solved exactly for $\eta\geq\eta_c=(|J_1'|-J_1)/J_1\approx 0.747$. The procedure of constructing the exact ground state wave function follows the spirit of the Majumdar-Ghosh (MG) point for the antiferromagnetic Heisenberg chain: when $J_1=J_1'=2J_2$, its ground state is a direct product of singlet dimers with twofold degeneracy~\cite{Majumdar1969}. The MG exact solution has been extended to the more general case of $J_1\neq J_1', J_1'=2J_2$ with exchange anisotropy $\eta$ for all $J>0$ by Shastry and Sutherland~\cite{Shastry1981}, and to cases with ferromagnetic exchange by Kanter (for a different model where not all NNN interactions are included)~\cite{Kanter}. We find that the technique can be applied to the zigzag Hamiltonian $H$ here and the ground state is also a direct product of singlet dimers on $J_1$ bonds.

The main steps of the solution are as follows.
First, for $J_1'=2J_2<0$, the product state of spin singlet $(\vert\uparrow\downarrow\rangle-\vert\downarrow\uparrow\rangle)/\sqrt{2}$ on all $J_1$ bonds, represented by thick solid lines in Fig. 2, for all the $J_1$ bond can be shown to be an eigenstate of $H$ for any $\eta\in [0,1]$ with eigenvalue $E_{\rm{eg}}=-M(2+\eta)J_1/4$, where $M$ is the number of triangles. Second, the total Hamiltonian is decomposed into sum of triangle Hamiltonians, $H=\sum_\ell h_\ell$, where $h_\ell$ is the Hamiltonian for a single triangle labeled by $\ell$. The ground state energy $e_\ell$ for $h_\ell$ can be calculated since it only involves three spins. Note that $M e_\ell$ serves as the lower bound of variational ground state energy. We find that for $\eta\geq 0.747$, $E_{\rm{eg}}=M e_\ell$, i.e. $E_{\rm{eg}}$ saturates the lower bound. Therefore, {\it the singlet product state must be the exact ground state}.
Interestingly, this product state of singlet dimers is smoothly deformed to the Haldane dimer phase at $\theta=0$, which can be understood from emergent spin-1 degree of freedom driven by strong ferromagnetic NN couplings~\cite{Furukawa2012}. Our model explicitly verifies the connection between these two cases, conjectured earlier using bosonization~\cite{Furukawa2012}.
Similarly, we find the ground state for the point $\eta=0$, $\theta\sim 42.4^\circ$ is the product of spin triplet $(\vert\uparrow\downarrow\rangle+\vert\downarrow\uparrow\rangle)/\sqrt{2}$ on all $J'_1$ bonds,  shown by the thick dashed lines in Fig.~2. Any ground state within the ED phase can be continuously deformed to this triplet product state without closing the gap. Similar to the point $\eta=1, \theta\sim 50.9^\circ$ shown in Fig.~\ref{fig:Hei40}(b),  the entanglement entropy $S^{\rm vN}$ of all the exact solvable cases are zero. Details on the exact solution 
can be found in Ref.~\cite{noteS}.

Both exact wave functions feature short range entanglement and preserve the symmetry of the Hamiltonian.
Both imply edge states: as the singlet or triplet valence bond is cut open at the edge, free ``dangling" spin-1/2 edge excitations are created,
similar to the Affleck-Kennedy-Lieb-Tasaki state \cite{affleck1987rigorous}. Each edge state is twofold degenerate and protected by, e.g., TR symmetry. Despite having the same symmetry, the SD and ED phases are topologically distinct. They cannot be deformed smoothly into each other if TR, inversion and $D_2$ symmetry of spin rotation $\pi$ about the $x$, $y$, and $z$ axis remain unbroken \cite{chen2011complete,pollmann2012symmetry,Ueda2014Chiral2}. Details on all open chain cases can be found in Ref.~\cite{noteS}.
The SD (ED) phase here is adiabatically connected to the $D_+$ ($D_-$) phase of $J_1-J_2$ model studied in Ref. \cite{Ueda2014Chiral2} for $J_1/J_2\in(-2.7,-1.5)$ and small bond alteration $\delta$. The $Z_2$ indices $\alpha,\beta,\gamma,\omega$ of these two SPT phases are tabulated in Ref. \cite{Ueda2014Chiral2}. Both SPT phases
 feature a double degeneracy in the entanglement spectrum \cite{pollmann2012symmetry}, and this is confirmed by our iTEBD calculation.

\textit{Experimental signatures.---} A first step toward realizing Hamiltonian Eq.~\eqref{eq:model} is to load polar molecules~\cite{Yan:2013xe} into a deep zigzag lattice \cite{Jo2015,PhysRevA.94.063632} with filling close to one. 
The SPT phases can be detected by measuring the edge excitations or the string order parameter. An open edge can be engineered by a strong local optical potential to terminate the zigzag chain or by creating local vacancies. Such control and probe seem within the reach of recently proposed site-resolved microscopy and spin-resolved detection for polar molecules~\cite{Covey_2018}. Then microwave spectroscopy may resolve edge states as a peak at ``forbidden energies" within the bulk gap. Furthermore, local perturbations can be applied to lift the edge degeneracy as outlined in Ref.~\cite{manmana2013topological}. Measurements of string order parameters have been achieved in a few systems~\cite{Hidden2006,Endres2013,Hidden2017,Hidden2018}. 

In summary, we have shown the zig-zag $XXZ$ model inspired by molecular gas experiments provides a promising platform for realizing SPT phases for spin-1/2 systems.
It unifies previous results in the Heisenberg limit by revealing the 
connections between them, and elucidates the nature of two robust SPT phases by finding their exact
ground states as product of singlet or triplet dimers. From this perspective, searching for and understanding the myriad of SPT phases could benefit from deforming the Hamiltonian to special anchor points where
the ground state wave function simplifies, as demonstrated here by exploiting the underlying triangular motif.
Other SPT phases in 1D can be potentially represented by such anchor points where their nature is intuitive and apparent from the exact wave functions. Finally, tuning the zig-zag angle $\gamma$ away from $30^\circ$ opens up a large parameter space of exchange couplings and the possibility of new SPT phases that deserve future investigation.

\acknowledgments
We thank Meng Cheng and Susan Yelin for helpful discussions. This work is supported by National Natural Science Foundation of China Grant No. 11804221 (H. Z.), Science and Technology Commission of Shanghai Municipality Grant No. 16DZ2260200 (H. Z. and W. V. L), AFOSR Grant No. FA9550-16-1-0006 (E. Z. and W. V. L.), NSF Grant No. PHY-1707484 (E. Z.) and MURI-ARO Grant No. W911NF-17-1-0323, ARO Grant No. W911NF-11-1-0230, and the Overseas Scholar Collaborative Program of NSF of China Grant No. 11429402 sponsored by Peking University (W. V. L.). X. W. G. is partially supported by the key NSFC Grant No.\ 11534014 and the National Key R\&D Program of China Grant No. 2017YFA0304500.


\begin{thebibliography}{41}%
\makeatletter
\providecommand \@ifxundefined [1]{%
 \@ifx{#1\undefined}
}%
\providecommand \@ifnum [1]{%
 \ifnum #1\expandafter \@firstoftwo
 \else \expandafter \@secondoftwo
 \fi
}%
\providecommand \@ifx [1]{%
 \ifx #1\expandafter \@firstoftwo
 \else \expandafter \@secondoftwo
 \fi
}%
\providecommand \natexlab [1]{#1}%
\providecommand \enquote  [1]{``#1''}%
\providecommand \bibnamefont  [1]{#1}%
\providecommand \bibfnamefont [1]{#1}%
\providecommand \citenamefont [1]{#1}%
\providecommand \href@noop [0]{\@secondoftwo}%
\providecommand \href [0]{\begingroup \@sanitize@url \@href}%
\providecommand \@href[1]{\@@startlink{#1}\@@href}%
\providecommand \@@href[1]{\endgroup#1\@@endlink}%
\providecommand \@sanitize@url [0]{\catcode `\\12\catcode `\$12\catcode
  `\&12\catcode `\#12\catcode `\^12\catcode `\_12\catcode `\%12\relax}%
\providecommand \@@startlink[1]{}%
\providecommand \@@endlink[0]{}%
\providecommand \url  [0]{\begingroup\@sanitize@url \@url }%
\providecommand \@url [1]{\endgroup\@href {#1}{\urlprefix }}%
\providecommand \urlprefix  [0]{URL }%
\providecommand \Eprint [0]{\href }%
\providecommand \doibase [0]{http://dx.doi.org/}%
\providecommand \selectlanguage [0]{\@gobble}%
\providecommand \bibinfo  [0]{\@secondoftwo}%
\providecommand \bibfield  [0]{\@secondoftwo}%
\providecommand \translation [1]{[#1]}%
\providecommand \BibitemOpen [0]{}%
\providecommand \bibitemStop [0]{}%
\providecommand \bibitemNoStop [0]{.\EOS\space}%
\providecommand \EOS [0]{\spacefactor3000\relax}%
\providecommand \BibitemShut  [1]{\csname bibitem#1\endcsname}%
\let\auto@bib@innerbib\@empty
\bibitem [{\citenamefont {Chen}\ \emph {et~al.}(2011)\citenamefont {Chen},
  \citenamefont {Gu},\ and\ \citenamefont {Wen}}]{chen2011complete}%
  \BibitemOpen
  \bibfield  {author} {\bibinfo {author} {\bibfnamefont {X.}~\bibnamefont
  {Chen}}, \bibinfo {author} {\bibfnamefont {Z.-C.}\ \bibnamefont {Gu}}, \ and\
  \bibinfo {author} {\bibfnamefont {X.-G.}\ \bibnamefont {Wen}},\ }\href@noop
  {} {\bibfield  {journal} {\bibinfo  {journal} {Physical Review B}\ }\textbf
  {\bibinfo {volume} {84}},\ \bibinfo {pages} {235128} (\bibinfo {year}
  {2011})}\BibitemShut {NoStop}%
\bibitem [{\citenamefont {Chen}\ \emph {et~al.}(2013)\citenamefont {Chen},
  \citenamefont {Gu}, \citenamefont {Liu},\ and\ \citenamefont
  {Wen}}]{chen2013symmetry}%
  \BibitemOpen
  \bibfield  {author} {\bibinfo {author} {\bibfnamefont {X.}~\bibnamefont
  {Chen}}, \bibinfo {author} {\bibfnamefont {Z.-C.}\ \bibnamefont {Gu}},
  \bibinfo {author} {\bibfnamefont {Z.-X.}\ \bibnamefont {Liu}}, \ and\
  \bibinfo {author} {\bibfnamefont {X.-G.}\ \bibnamefont {Wen}},\ }\href@noop
  {} {\bibfield  {journal} {\bibinfo  {journal} {Physical Review B}\ }\textbf
  {\bibinfo {volume} {87}},\ \bibinfo {pages} {155114} (\bibinfo {year}
  {2013})}\BibitemShut {NoStop}%
\bibitem [{\citenamefont {Senthil}(2015)}]{senthil2015symmetry}%
  \BibitemOpen
  \bibfield  {author} {\bibinfo {author} {\bibfnamefont {T.}~\bibnamefont
  {Senthil}},\ }\href@noop {} {\bibfield  {journal} {\bibinfo  {journal} {Annu.
  Rev. Condens. Matter Phys.}\ }\textbf {\bibinfo {volume} {6}},\ \bibinfo
  {pages} {299} (\bibinfo {year} {2015})}\BibitemShut {NoStop}%
\bibitem [{\citenamefont {Haldane}(1983)}]{Haldane}%
  \BibitemOpen
  \bibfield  {author} {\bibinfo {author} {\bibfnamefont {F.~D.~M.}\
  \bibnamefont {Haldane}},\ }\href {\doibase 10.1103/PhysRevLett.50.1153}
  {\bibfield  {journal} {\bibinfo  {journal} {Phys. Rev. Lett.}\ }\textbf
  {\bibinfo {volume} {50}},\ \bibinfo {pages} {1153} (\bibinfo {year}
  {1983})}\BibitemShut {NoStop}%
\bibitem [{\citenamefont {Furukawa}\ \emph {et~al.}(2012)\citenamefont
  {Furukawa}, \citenamefont {Sato}, \citenamefont {Onoda},\ and\ \citenamefont
  {Furusaki}}]{Furukawa2012}%
  \BibitemOpen
  \bibfield  {author} {\bibinfo {author} {\bibfnamefont {S.}~\bibnamefont
  {Furukawa}}, \bibinfo {author} {\bibfnamefont {M.}~\bibnamefont {Sato}},
  \bibinfo {author} {\bibfnamefont {S.}~\bibnamefont {Onoda}}, \ and\ \bibinfo
  {author} {\bibfnamefont {A.}~\bibnamefont {Furusaki}},\ }\href {\doibase
  10.1103/PhysRevB.86.094417} {\bibfield  {journal} {\bibinfo  {journal} {Phys.
  Rev. B}\ }\textbf {\bibinfo {volume} {86}},\ \bibinfo {pages} {094417}
  (\bibinfo {year} {2012})}\BibitemShut {NoStop}%
\bibitem [{\citenamefont {White}(1996)}]{White1}%
  \BibitemOpen
  \bibfield  {author} {\bibinfo {author} {\bibfnamefont {S.~R.}\ \bibnamefont
  {White}},\ }\href {\doibase 10.1103/PhysRevB.53.52} {\bibfield  {journal}
  {\bibinfo  {journal} {Phys. Rev. B}\ }\textbf {\bibinfo {volume} {53}},\
  \bibinfo {pages} {52} (\bibinfo {year} {1996})}\BibitemShut {NoStop}%
\bibitem [{\citenamefont {Hida}(1992)}]{alterchain}%
  \BibitemOpen
  \bibfield  {author} {\bibinfo {author} {\bibfnamefont {K.}~\bibnamefont
  {Hida}},\ }\href {\doibase 10.1103/PhysRevB.45.2207} {\bibfield  {journal}
  {\bibinfo  {journal} {Phys. Rev. B}\ }\textbf {\bibinfo {volume} {45}},\
  \bibinfo {pages} {2207} (\bibinfo {year} {1992})}\BibitemShut {NoStop}%
\bibitem [{\citenamefont {Kanter}(1989)}]{Kanter}%
  \BibitemOpen
  \bibfield  {author} {\bibinfo {author} {\bibfnamefont {I.}~\bibnamefont
  {Kanter}},\ }\href {\doibase 10.1103/PhysRevB.39.7270} {\bibfield  {journal}
  {\bibinfo  {journal} {Phys. Rev. B}\ }\textbf {\bibinfo {volume} {39}},\
  \bibinfo {pages} {7270} (\bibinfo {year} {1989})}\BibitemShut {NoStop}%
\bibitem [{\citenamefont {Vekua}\ \emph {et~al.}(2003)\citenamefont {Vekua},
  \citenamefont {Japaridze},\ and\ \citenamefont {Mikeska}}]{FMleg}%
  \BibitemOpen
  \bibfield  {author} {\bibinfo {author} {\bibfnamefont {T.}~\bibnamefont
  {Vekua}}, \bibinfo {author} {\bibfnamefont {G.~I.}\ \bibnamefont
  {Japaridze}}, \ and\ \bibinfo {author} {\bibfnamefont {H.-J.}\ \bibnamefont
  {Mikeska}},\ }\href {\doibase 10.1103/PhysRevB.67.064419} {\bibfield
  {journal} {\bibinfo  {journal} {Phys. Rev. B}\ }\textbf {\bibinfo {volume}
  {67}},\ \bibinfo {pages} {064419} (\bibinfo {year} {2003})}\BibitemShut
  {NoStop}%
\bibitem [{\citenamefont {Ueda}\ and\ \citenamefont
  {Onoda}(2014{\natexlab{a}})}]{Ueda2014Chiral2}%
  \BibitemOpen
  \bibfield  {author} {\bibinfo {author} {\bibfnamefont {H.}~\bibnamefont
  {Ueda}}\ and\ \bibinfo {author} {\bibfnamefont {S.}~\bibnamefont {Onoda}},\
  }\href {\doibase 10.1103/PhysRevB.90.214425} {\bibfield  {journal} {\bibinfo
  {journal} {Phys. Rev. B}\ }\textbf {\bibinfo {volume} {90}},\ \bibinfo
  {pages} {214425} (\bibinfo {year} {2014}{\natexlab{a}})}\BibitemShut
  {NoStop}%
\bibitem [{\citenamefont {Liu}\ \emph {et~al.}(2012)\citenamefont {Liu},
  \citenamefont {Yang}, \citenamefont {Han}, \citenamefont {Yi},\ and\
  \citenamefont {Wen}}]{liu2012symmetry}%
  \BibitemOpen
  \bibfield  {author} {\bibinfo {author} {\bibfnamefont {Z.-X.}\ \bibnamefont
  {Liu}}, \bibinfo {author} {\bibfnamefont {Z.-B.}\ \bibnamefont {Yang}},
  \bibinfo {author} {\bibfnamefont {Y.-J.}\ \bibnamefont {Han}}, \bibinfo
  {author} {\bibfnamefont {W.}~\bibnamefont {Yi}}, \ and\ \bibinfo {author}
  {\bibfnamefont {X.-G.}\ \bibnamefont {Wen}},\ }\href@noop {} {\bibfield
  {journal} {\bibinfo  {journal} {Physical Review B}\ }\textbf {\bibinfo
  {volume} {86}},\ \bibinfo {pages} {195122} (\bibinfo {year}
  {2012})}\BibitemShut {NoStop}%
\bibitem [{\citenamefont {Manmana}\ \emph {et~al.}(2013)\citenamefont
  {Manmana}, \citenamefont {Stoudenmire}, \citenamefont {Hazzard},
  \citenamefont {Rey},\ and\ \citenamefont
  {Gorshkov}}]{manmana2013topological}%
  \BibitemOpen
  \bibfield  {author} {\bibinfo {author} {\bibfnamefont {S.~R.}\ \bibnamefont
  {Manmana}}, \bibinfo {author} {\bibfnamefont {E.}~\bibnamefont
  {Stoudenmire}}, \bibinfo {author} {\bibfnamefont {K.~R.}\ \bibnamefont
  {Hazzard}}, \bibinfo {author} {\bibfnamefont {A.~M.}\ \bibnamefont {Rey}}, \
  and\ \bibinfo {author} {\bibfnamefont {A.~V.}\ \bibnamefont {Gorshkov}},\
  }\href@noop {} {\bibfield  {journal} {\bibinfo  {journal} {Physical Review
  B}\ }\textbf {\bibinfo {volume} {87}},\ \bibinfo {pages} {081106} (\bibinfo
  {year} {2013})}\BibitemShut {NoStop}%
\bibitem [{\citenamefont {Song}\ \emph {et~al.}(2018)\citenamefont {Song},
  \citenamefont {Zhang}, \citenamefont {He}, \citenamefont {Poon},
  \citenamefont {Hajiyev}, \citenamefont {Zhang}, \citenamefont {Liu},\ and\
  \citenamefont {Jo}}]{bSPT2018}%
  \BibitemOpen
  \bibfield  {author} {\bibinfo {author} {\bibfnamefont {B.}~\bibnamefont
  {Song}}, \bibinfo {author} {\bibfnamefont {L.}~\bibnamefont {Zhang}},
  \bibinfo {author} {\bibfnamefont {C.}~\bibnamefont {He}}, \bibinfo {author}
  {\bibfnamefont {T.~F.~J.}\ \bibnamefont {Poon}}, \bibinfo {author}
  {\bibfnamefont {E.}~\bibnamefont {Hajiyev}}, \bibinfo {author} {\bibfnamefont
  {S.}~\bibnamefont {Zhang}}, \bibinfo {author} {\bibfnamefont {X.-J.}\
  \bibnamefont {Liu}}, \ and\ \bibinfo {author} {\bibfnamefont {G.-B.}\
  \bibnamefont {Jo}},\ }\href
  {http://advances.sciencemag.org/content/4/2/eaao4748} {\bibfield  {journal}
  {\bibinfo  {journal} {Science Advances}\ }\textbf {\bibinfo {volume} {4}},\
  \bibinfo {pages} {2, eaao4748} (\bibinfo {year} {2018})}\BibitemShut
  {NoStop}%
\bibitem [{\citenamefont {de~L\'es\'eleuc}\ \emph {et~al.}(2018)\citenamefont
  {de~L\'es\'eleuc}, \citenamefont {Lienhard}, \citenamefont {Scholl},
  \citenamefont {Barredo}, \citenamefont {Weber}, \citenamefont {Lang},
  \citenamefont {B\"uchler}, \citenamefont {Lahaye},\ and\ \citenamefont
  {Browaeys}}]{fSPTarxiv2018}%
  \BibitemOpen
  \bibfield  {author} {\bibinfo {author} {\bibfnamefont {S.}~\bibnamefont
  {de~L\'es\'eleuc}}, \bibinfo {author} {\bibfnamefont {V.}~\bibnamefont
  {Lienhard}}, \bibinfo {author} {\bibfnamefont {P.}~\bibnamefont {Scholl}},
  \bibinfo {author} {\bibfnamefont {D.}~\bibnamefont {Barredo}}, \bibinfo
  {author} {\bibfnamefont {S.}~\bibnamefont {Weber}}, \bibinfo {author}
  {\bibfnamefont {N.}~\bibnamefont {Lang}}, \bibinfo {author} {\bibfnamefont
  {H.~P.}\ \bibnamefont {B\"uchler}}, \bibinfo {author} {\bibfnamefont
  {T.}~\bibnamefont {Lahaye}}, \ and\ \bibinfo {author} {\bibfnamefont
  {A.}~\bibnamefont {Browaeys}},\ }\href@noop {} {} (\bibinfo {year} {2018}),\
  \Eprint {http://arxiv.org/abs/arXiv:1810.13286} {arXiv:1810.13286}
  \BibitemShut {NoStop}%
\bibitem [{\citenamefont {Yan}\ \emph {et~al.}(2013)\citenamefont {Yan},
  \citenamefont {Moses}, \citenamefont {Gadway}, \citenamefont {Covey},
  \citenamefont {Hazzard}, \citenamefont {Rey}, \citenamefont {Jin},\ and\
  \citenamefont {Ye}}]{Yan:2013xe}%
  \BibitemOpen
  \bibfield  {author} {\bibinfo {author} {\bibfnamefont {B.}~\bibnamefont
  {Yan}}, \bibinfo {author} {\bibfnamefont {S.~A.}\ \bibnamefont {Moses}},
  \bibinfo {author} {\bibfnamefont {B.}~\bibnamefont {Gadway}}, \bibinfo
  {author} {\bibfnamefont {J.~P.}\ \bibnamefont {Covey}}, \bibinfo {author}
  {\bibfnamefont {K.~R.~A.}\ \bibnamefont {Hazzard}}, \bibinfo {author}
  {\bibfnamefont {A.~M.}\ \bibnamefont {Rey}}, \bibinfo {author} {\bibfnamefont
  {D.~S.}\ \bibnamefont {Jin}}, \ and\ \bibinfo {author} {\bibfnamefont
  {J.}~\bibnamefont {Ye}},\ }\href@noop {} {\bibfield  {journal} {\bibinfo
  {journal} {Nature}\ }\textbf {\bibinfo {volume} {501}},\ \bibinfo {pages}
  {521} (\bibinfo {year} {2013})}\BibitemShut {NoStop}%
\bibitem [{\citenamefont {Hazzard}\ \emph {et~al.}(2014)\citenamefont
  {Hazzard}, \citenamefont {Gadway}, \citenamefont {Foss-Feig}, \citenamefont
  {Yan}, \citenamefont {Moses}, \citenamefont {Covey}, \citenamefont {Yao},
  \citenamefont {Lukin}, \citenamefont {Ye}, \citenamefont {Jin},\ and\
  \citenamefont {Rey}}]{PhysRevLett.113.195302}%
  \BibitemOpen
  \bibfield  {author} {\bibinfo {author} {\bibfnamefont {K.~R.~A.}\
  \bibnamefont {Hazzard}}, \bibinfo {author} {\bibfnamefont {B.}~\bibnamefont
  {Gadway}}, \bibinfo {author} {\bibfnamefont {M.}~\bibnamefont {Foss-Feig}},
  \bibinfo {author} {\bibfnamefont {B.}~\bibnamefont {Yan}}, \bibinfo {author}
  {\bibfnamefont {S.~A.}\ \bibnamefont {Moses}}, \bibinfo {author}
  {\bibfnamefont {J.~P.}\ \bibnamefont {Covey}}, \bibinfo {author}
  {\bibfnamefont {N.~Y.}\ \bibnamefont {Yao}}, \bibinfo {author} {\bibfnamefont
  {M.~D.}\ \bibnamefont {Lukin}}, \bibinfo {author} {\bibfnamefont
  {J.}~\bibnamefont {Ye}}, \bibinfo {author} {\bibfnamefont {D.~S.}\
  \bibnamefont {Jin}}, \ and\ \bibinfo {author} {\bibfnamefont {A.~M.}\
  \bibnamefont {Rey}},\ }\href {\doibase 10.1103/PhysRevLett.113.195302}
  {\bibfield  {journal} {\bibinfo  {journal} {Phys. Rev. Lett.}\ }\textbf
  {\bibinfo {volume} {113}},\ \bibinfo {pages} {195302} (\bibinfo {year}
  {2014})}\BibitemShut {NoStop}%
\bibitem [{\citenamefont {Gorshkov}\ \emph {et~al.}(2011)\citenamefont
  {Gorshkov}, \citenamefont {Manmana}, \citenamefont {Chen}, \citenamefont
  {Ye}, \citenamefont {Demler}, \citenamefont {Lukin},\ and\ \citenamefont
  {Rey}}]{PhysRevLett.107.115301}%
  \BibitemOpen
  \bibfield  {author} {\bibinfo {author} {\bibfnamefont {A.~V.}\ \bibnamefont
  {Gorshkov}}, \bibinfo {author} {\bibfnamefont {S.~R.}\ \bibnamefont
  {Manmana}}, \bibinfo {author} {\bibfnamefont {G.}~\bibnamefont {Chen}},
  \bibinfo {author} {\bibfnamefont {J.}~\bibnamefont {Ye}}, \bibinfo {author}
  {\bibfnamefont {E.}~\bibnamefont {Demler}}, \bibinfo {author} {\bibfnamefont
  {M.~D.}\ \bibnamefont {Lukin}}, \ and\ \bibinfo {author} {\bibfnamefont
  {A.~M.}\ \bibnamefont {Rey}},\ }\href {\doibase
  10.1103/PhysRevLett.107.115301} {\bibfield  {journal} {\bibinfo  {journal}
  {Phys. Rev. Lett.}\ }\textbf {\bibinfo {volume} {107}},\ \bibinfo {pages}
  {115301} (\bibinfo {year} {2011})}\BibitemShut {NoStop}%
\bibitem [{\citenamefont {de~Paz}\ \emph {et~al.}(2013)\citenamefont {de~Paz},
  \citenamefont {Sharma}, \citenamefont {Chotia}, \citenamefont {Mar\'echal},
  \citenamefont {Huckans}, \citenamefont {Pedri}, \citenamefont {Santos},
  \citenamefont {Gorceix}, \citenamefont {Vernac},\ and\ \citenamefont
  {Laburthe-Tolra}}]{PhysRevLett.111.185305}%
  \BibitemOpen
  \bibfield  {author} {\bibinfo {author} {\bibfnamefont {A.}~\bibnamefont
  {de~Paz}}, \bibinfo {author} {\bibfnamefont {A.}~\bibnamefont {Sharma}},
  \bibinfo {author} {\bibfnamefont {A.}~\bibnamefont {Chotia}}, \bibinfo
  {author} {\bibfnamefont {E.}~\bibnamefont {Mar\'echal}}, \bibinfo {author}
  {\bibfnamefont {J.~H.}\ \bibnamefont {Huckans}}, \bibinfo {author}
  {\bibfnamefont {P.}~\bibnamefont {Pedri}}, \bibinfo {author} {\bibfnamefont
  {L.}~\bibnamefont {Santos}}, \bibinfo {author} {\bibfnamefont
  {O.}~\bibnamefont {Gorceix}}, \bibinfo {author} {\bibfnamefont
  {L.}~\bibnamefont {Vernac}}, \ and\ \bibinfo {author} {\bibfnamefont
  {B.}~\bibnamefont {Laburthe-Tolra}},\ }\href {\doibase
  10.1103/PhysRevLett.111.185305} {\bibfield  {journal} {\bibinfo  {journal}
  {Phys. Rev. Lett.}\ }\textbf {\bibinfo {volume} {111}},\ \bibinfo {pages}
  {185305} (\bibinfo {year} {2013})}\BibitemShut {NoStop}%
\bibitem [{\citenamefont {Yao}\ \emph {et~al.}(2018)\citenamefont {Yao},
  \citenamefont {Zaletel}, \citenamefont {Stamper-Kurn},\ and\ \citenamefont
  {Vishwanath}}]{DSL2015}%
  \BibitemOpen
  \bibfield  {author} {\bibinfo {author} {\bibfnamefont {N.~Y.}\ \bibnamefont
  {Yao}}, \bibinfo {author} {\bibfnamefont {M.~P.}\ \bibnamefont {Zaletel}},
  \bibinfo {author} {\bibfnamefont {D.~M.}\ \bibnamefont {Stamper-Kurn}}, \
  and\ \bibinfo {author} {\bibfnamefont {A.}~\bibnamefont {Vishwanath}},\
  }\href {\doibase 10.1038/s41567-017-0030-7} {\bibfield  {journal} {\bibinfo
  {journal} {Nature Physics}\ }\textbf {\bibinfo {volume} {14}},\ \bibinfo
  {pages} {405} (\bibinfo {year} {2018})}\BibitemShut {NoStop}%
\bibitem [{\citenamefont {Zou}\ \emph {et~al.}(2017)\citenamefont {Zou},
  \citenamefont {Zhao},\ and\ \citenamefont {Liu}}]{Our2017}%
  \BibitemOpen
  \bibfield  {author} {\bibinfo {author} {\bibfnamefont {H.}~\bibnamefont
  {Zou}}, \bibinfo {author} {\bibfnamefont {E.}~\bibnamefont {Zhao}}, \ and\
  \bibinfo {author} {\bibfnamefont {W.~V.}\ \bibnamefont {Liu}},\ }\href
  {\doibase 10.1103/PhysRevLett.119.050401} {\bibfield  {journal} {\bibinfo
  {journal} {Phys. Rev. Lett.}\ }\textbf {\bibinfo {volume} {119}},\ \bibinfo
  {pages} {050401} (\bibinfo {year} {2017})}\BibitemShut {NoStop}%
\bibitem [{\citenamefont {Kele\ifmmode~\mbox{\c{s}}\else \c{s}\fi{}}\ and\
  \citenamefont {Zhao}(2018{\natexlab{a}})}]{keles2018absence}%
  \BibitemOpen
  \bibfield  {author} {\bibinfo {author} {\bibfnamefont {A.}~\bibnamefont
  {Kele\ifmmode~\mbox{\c{s}}\else \c{s}\fi{}}}\ and\ \bibinfo {author}
  {\bibfnamefont {E.}~\bibnamefont {Zhao}},\ }\href {\doibase
  10.1103/PhysRevLett.120.187202} {\bibfield  {journal} {\bibinfo  {journal}
  {Phys. Rev. Lett.}\ }\textbf {\bibinfo {volume} {120}},\ \bibinfo {pages}
  {187202} (\bibinfo {year} {2018}{\natexlab{a}})}\BibitemShut {NoStop}%
\bibitem [{\citenamefont {Kele\ifmmode~\mbox{\c{s}}\else \c{s}\fi{}}\ and\
  \citenamefont {Zhao}(2018{\natexlab{b}})}]{keles-prb}%
  \BibitemOpen
  \bibfield  {author} {\bibinfo {author} {\bibfnamefont {A.}~\bibnamefont
  {Kele\ifmmode~\mbox{\c{s}}\else \c{s}\fi{}}}\ and\ \bibinfo {author}
  {\bibfnamefont {E.}~\bibnamefont {Zhao}},\ }\href {\doibase
  10.1103/PhysRevB.97.245105} {\bibfield  {journal} {\bibinfo  {journal} {Phys.
  Rev. B}\ }\textbf {\bibinfo {volume} {97}},\ \bibinfo {pages} {245105}
  (\bibinfo {year} {2018}{\natexlab{b}})}\BibitemShut {NoStop}%
\bibitem [{\citenamefont {Hikihara}\ \emph {et~al.}(2001)\citenamefont
  {Hikihara}, \citenamefont {Kaburagi},\ and\ \citenamefont
  {Kawamura}}]{PhysRevB.63.174430}%
  \BibitemOpen
  \bibfield  {author} {\bibinfo {author} {\bibfnamefont {T.}~\bibnamefont
  {Hikihara}}, \bibinfo {author} {\bibfnamefont {M.}~\bibnamefont {Kaburagi}},
  \ and\ \bibinfo {author} {\bibfnamefont {H.}~\bibnamefont {Kawamura}},\
  }\href {\doibase 10.1103/PhysRevB.63.174430} {\bibfield  {journal} {\bibinfo
  {journal} {Phys. Rev. B}\ }\textbf {\bibinfo {volume} {63}},\ \bibinfo
  {pages} {174430} (\bibinfo {year} {2001})}\BibitemShut {NoStop}%
\bibitem [{\citenamefont {Ueda}\ and\ \citenamefont
  {Onoda}(2014{\natexlab{b}})}]{Ueda2014Chiral}%
  \BibitemOpen
  \bibfield  {author} {\bibinfo {author} {\bibfnamefont {H.}~\bibnamefont
  {Ueda}}\ and\ \bibinfo {author} {\bibfnamefont {S.}~\bibnamefont {Onoda}},\
  }\href {\doibase 10.1103/PhysRevB.89.024407} {\bibfield  {journal} {\bibinfo
  {journal} {Phys. Rev. B}\ }\textbf {\bibinfo {volume} {89}},\ \bibinfo
  {pages} {024407} (\bibinfo {year} {2014}{\natexlab{b}})}\BibitemShut
  {NoStop}%
\bibitem [{Note1()}]{Note1}%
  \BibitemOpen
  \bibinfo {note} {In Ref. \cite {Furukawa2012}, a schematic phase diagram
  (Fig. 11) was conjectured based on bosonization.}\BibitemShut {Stop}%
\bibitem [{\citenamefont {Wang}\ \emph {et~al.}(2017)\citenamefont {Wang},
  \citenamefont {Otterbach},\ and\ \citenamefont {Yelin}}]{Yelin2017}%
  \BibitemOpen
  \bibfield  {author} {\bibinfo {author} {\bibfnamefont {Q.}~\bibnamefont
  {Wang}}, \bibinfo {author} {\bibfnamefont {J.}~\bibnamefont {Otterbach}}, \
  and\ \bibinfo {author} {\bibfnamefont {S.~F.}\ \bibnamefont {Yelin}},\ }\href
  {\doibase 10.1103/PhysRevA.96.043615} {\bibfield  {journal} {\bibinfo
  {journal} {Phys. Rev. A}\ }\textbf {\bibinfo {volume} {96}},\ \bibinfo
  {pages} {043615} (\bibinfo {year} {2017})}\BibitemShut {NoStop}%
\bibitem [{\citenamefont {Greschner}\ \emph {et~al.}(2013)\citenamefont
  {Greschner}, \citenamefont {Santos},\ and\ \citenamefont
  {Vekua}}]{Vekua2013}%
  \BibitemOpen
  \bibfield  {author} {\bibinfo {author} {\bibfnamefont {S.}~\bibnamefont
  {Greschner}}, \bibinfo {author} {\bibfnamefont {L.}~\bibnamefont {Santos}}, \
  and\ \bibinfo {author} {\bibfnamefont {T.}~\bibnamefont {Vekua}},\ }\href
  {\doibase 10.1103/PhysRevA.87.033609} {\bibfield  {journal} {\bibinfo
  {journal} {Phys. Rev. A}\ }\textbf {\bibinfo {volume} {87}},\ \bibinfo
  {pages} {033609} (\bibinfo {year} {2013})}\BibitemShut {NoStop}%
\bibitem [{\citenamefont {Zhang}\ and\ \citenamefont {Jo}(2015)}]{Jo2015}%
  \BibitemOpen
  \bibfield  {author} {\bibinfo {author} {\bibfnamefont {T.}~\bibnamefont
  {Zhang}}\ and\ \bibinfo {author} {\bibfnamefont {G.-B.}\ \bibnamefont {Jo}},\
  }\href {http://dx.doi.org/10.1038/srep16044} {\bibfield  {journal} {\bibinfo
  {journal} {Scientific Reports}\ }\textbf {\bibinfo {volume} {5}},\ \bibinfo
  {pages} {16044 EP } (\bibinfo {year} {2015})}\BibitemShut {NoStop}%
\bibitem [{\citenamefont {Vidal}(2007)}]{iTEBD}%
  \BibitemOpen
  \bibfield  {author} {\bibinfo {author} {\bibfnamefont {G.}~\bibnamefont
  {Vidal}},\ }\href {\doibase 10.1103/PhysRevLett.98.070201} {\bibfield
  {journal} {\bibinfo  {journal} {Phys. Rev. Lett.}\ }\textbf {\bibinfo
  {volume} {98}},\ \bibinfo {pages} {070201} (\bibinfo {year}
  {2007})}\BibitemShut {NoStop}%
\bibitem [{not()}]{noteS}%
  \BibitemOpen
  \href@noop {} {}\bibinfo {note} {See Supplemental Material for detail
  description of exact solution, iTEBD calculation, and
  bosonization.}\BibitemShut {Stop}%
\bibitem [{\citenamefont {den Nijs}\ and\ \citenamefont
  {Rommelse}(1989)}]{string1989}%
  \BibitemOpen
  \bibfield  {author} {\bibinfo {author} {\bibfnamefont {M.}~\bibnamefont {den
  Nijs}}\ and\ \bibinfo {author} {\bibfnamefont {K.}~\bibnamefont {Rommelse}},\
  }\href@noop {} {\bibfield  {journal} {\bibinfo  {journal} {Phys. Rev. B}\
  }\textbf {\bibinfo {volume} {40}},\ \bibinfo {pages} {4709} (\bibinfo {year}
  {1989})}\BibitemShut {NoStop}%
\bibitem [{\citenamefont {Majumdar}\ and\ \citenamefont
  {Ghosh}(1969)}]{Majumdar1969}%
  \BibitemOpen
  \bibfield  {author} {\bibinfo {author} {\bibfnamefont {C.~K.}\ \bibnamefont
  {Majumdar}}\ and\ \bibinfo {author} {\bibfnamefont {D.~K.}\ \bibnamefont
  {Ghosh}},\ }\href {\doibase 10.1063/1.1664979} {\bibfield  {journal}
  {\bibinfo  {journal} {Journal of Mathematical Physics}\ }\textbf {\bibinfo
  {volume} {10}},\ \bibinfo {pages} {1399} (\bibinfo {year}
  {1969})}\BibitemShut {NoStop}%
\bibitem [{\citenamefont {Shastry}\ and\ \citenamefont
  {Sutherland}(1981)}]{Shastry1981}%
  \BibitemOpen
  \bibfield  {author} {\bibinfo {author} {\bibfnamefont {B.~S.}\ \bibnamefont
  {Shastry}}\ and\ \bibinfo {author} {\bibfnamefont {B.}~\bibnamefont
  {Sutherland}},\ }\href {\doibase 10.1103/PhysRevLett.47.964} {\bibfield
  {journal} {\bibinfo  {journal} {Phys. Rev. Lett.}\ }\textbf {\bibinfo
  {volume} {47}},\ \bibinfo {pages} {964} (\bibinfo {year} {1981})}\BibitemShut
  {NoStop}%
\bibitem [{\citenamefont {Affleck}\ \emph {et~al.}(1987)\citenamefont
  {Affleck}, \citenamefont {Kennedy}, \citenamefont {Lieb},\ and\ \citenamefont
  {Tasaki}}]{affleck1987rigorous}%
  \BibitemOpen
  \bibfield  {author} {\bibinfo {author} {\bibfnamefont {I.}~\bibnamefont
  {Affleck}}, \bibinfo {author} {\bibfnamefont {T.}~\bibnamefont {Kennedy}},
  \bibinfo {author} {\bibfnamefont {E.~H.}\ \bibnamefont {Lieb}}, \ and\
  \bibinfo {author} {\bibfnamefont {H.}~\bibnamefont {Tasaki}},\ }\href@noop {}
  {\bibfield  {journal} {\bibinfo  {journal} {Physical review letters}\
  }\textbf {\bibinfo {volume} {59}},\ \bibinfo {pages} {799} (\bibinfo {year}
  {1987})}\BibitemShut {NoStop}%
\bibitem [{\citenamefont {Pollmann}\ \emph {et~al.}(2012)\citenamefont
  {Pollmann}, \citenamefont {Berg}, \citenamefont {Turner},\ and\ \citenamefont
  {Oshikawa}}]{pollmann2012symmetry}%
  \BibitemOpen
  \bibfield  {author} {\bibinfo {author} {\bibfnamefont {F.}~\bibnamefont
  {Pollmann}}, \bibinfo {author} {\bibfnamefont {E.}~\bibnamefont {Berg}},
  \bibinfo {author} {\bibfnamefont {A.~M.}\ \bibnamefont {Turner}}, \ and\
  \bibinfo {author} {\bibfnamefont {M.}~\bibnamefont {Oshikawa}},\ }\href@noop
  {} {\bibfield  {journal} {\bibinfo  {journal} {Physical review b}\ }\textbf
  {\bibinfo {volume} {85}},\ \bibinfo {pages} {075125} (\bibinfo {year}
  {2012})}\BibitemShut {NoStop}%
\bibitem [{\citenamefont {Anisimovas}\ \emph {et~al.}(2016)\citenamefont
  {Anisimovas}, \citenamefont {Ra\ifmmode \check{c}\else
  \v{c}\fi{}i\ifmmode~\bar{u}\else \={u}\fi{}nas}, \citenamefont {Str\"ater},
  \citenamefont {Eckardt}, \citenamefont {Spielman},\ and\ \citenamefont
  {Juzeli\ifmmode~\bar{u}\else \={u}\fi{}nas}}]{PhysRevA.94.063632}%
  \BibitemOpen
  \bibfield  {author} {\bibinfo {author} {\bibfnamefont {E.}~\bibnamefont
  {Anisimovas}}, \bibinfo {author} {\bibfnamefont {M.}~\bibnamefont {Ra\ifmmode
  \check{c}\else \v{c}\fi{}i\ifmmode~\bar{u}\else \={u}\fi{}nas}}, \bibinfo
  {author} {\bibfnamefont {C.}~\bibnamefont {Str\"ater}}, \bibinfo {author}
  {\bibfnamefont {A.}~\bibnamefont {Eckardt}}, \bibinfo {author} {\bibfnamefont
  {I.~B.}\ \bibnamefont {Spielman}}, \ and\ \bibinfo {author} {\bibfnamefont
  {G.}~\bibnamefont {Juzeli\ifmmode~\bar{u}\else \={u}\fi{}nas}},\ }\href
  {\doibase 10.1103/PhysRevA.94.063632} {\bibfield  {journal} {\bibinfo
  {journal} {Phys. Rev. A}\ }\textbf {\bibinfo {volume} {94}},\ \bibinfo
  {pages} {063632} (\bibinfo {year} {2016})}\BibitemShut {NoStop}%
\bibitem [{\citenamefont {Covey}\ \emph {et~al.}(2018)\citenamefont {Covey},
  \citenamefont {Marco}, \citenamefont {Acevedo}, \citenamefont {Rey},\ and\
  \citenamefont {Ye}}]{Covey_2018}%
  \BibitemOpen
  \bibfield  {author} {\bibinfo {author} {\bibfnamefont {J.~P.}\ \bibnamefont
  {Covey}}, \bibinfo {author} {\bibfnamefont {L.~D.}\ \bibnamefont {Marco}},
  \bibinfo {author} {\bibfnamefont {{\'{O}}.~L.}\ \bibnamefont {Acevedo}},
  \bibinfo {author} {\bibfnamefont {A.~M.}\ \bibnamefont {Rey}}, \ and\
  \bibinfo {author} {\bibfnamefont {J.}~\bibnamefont {Ye}},\ }\href {\doibase
  10.1088/1367-2630/aaba65} {\bibfield  {journal} {\bibinfo  {journal} {New
  Journal of Physics}\ }\textbf {\bibinfo {volume} {20}},\ \bibinfo {pages}
  {043031} (\bibinfo {year} {2018})}\BibitemShut {NoStop}%
\bibitem [{\citenamefont {Dalla~Torre}\ \emph {et~al.}(2006)\citenamefont
  {Dalla~Torre}, \citenamefont {Berg},\ and\ \citenamefont
  {Altman}}]{Hidden2006}%
  \BibitemOpen
  \bibfield  {author} {\bibinfo {author} {\bibfnamefont {E.~G.}\ \bibnamefont
  {Dalla~Torre}}, \bibinfo {author} {\bibfnamefont {E.}~\bibnamefont {Berg}}, \
  and\ \bibinfo {author} {\bibfnamefont {E.}~\bibnamefont {Altman}},\ }\href
  {\doibase 10.1103/PhysRevLett.97.260401} {\bibfield  {journal} {\bibinfo
  {journal} {Phys. Rev. Lett.}\ }\textbf {\bibinfo {volume} {97}},\ \bibinfo
  {pages} {260401} (\bibinfo {year} {2006})}\BibitemShut {NoStop}%
\bibitem [{\citenamefont {Endres}\ \emph {et~al.}(2013)\citenamefont {Endres},
  \citenamefont {Cheneau}, \citenamefont {Fukuhara}, \citenamefont
  {Weitenberg}, \citenamefont {Schau{\ss}}, \citenamefont {Gross},
  \citenamefont {Mazza}, \citenamefont {Ba{\~{n}}uls}, \citenamefont {Pollet},
  \citenamefont {Bloch},\ and\ \citenamefont {Kuhr}}]{Endres2013}%
  \BibitemOpen
  \bibfield  {author} {\bibinfo {author} {\bibfnamefont {M.}~\bibnamefont
  {Endres}}, \bibinfo {author} {\bibfnamefont {M.}~\bibnamefont {Cheneau}},
  \bibinfo {author} {\bibfnamefont {T.}~\bibnamefont {Fukuhara}}, \bibinfo
  {author} {\bibfnamefont {C.}~\bibnamefont {Weitenberg}}, \bibinfo {author}
  {\bibfnamefont {P.}~\bibnamefont {Schau{\ss}}}, \bibinfo {author}
  {\bibfnamefont {C.}~\bibnamefont {Gross}}, \bibinfo {author} {\bibfnamefont
  {L.}~\bibnamefont {Mazza}}, \bibinfo {author} {\bibfnamefont {M.~C.}\
  \bibnamefont {Ba{\~{n}}uls}}, \bibinfo {author} {\bibfnamefont
  {L.}~\bibnamefont {Pollet}}, \bibinfo {author} {\bibfnamefont
  {I.}~\bibnamefont {Bloch}}, \ and\ \bibinfo {author} {\bibfnamefont
  {S.}~\bibnamefont {Kuhr}},\ }\href {\doibase 10.1007/s00340-013-5552-9}
  {\bibfield  {journal} {\bibinfo  {journal} {Applied Physics B}\ }\textbf
  {\bibinfo {volume} {113}},\ \bibinfo {pages} {27} (\bibinfo {year}
  {2013})}\BibitemShut {NoStop}%
\bibitem [{\citenamefont {Cardarelli}\ \emph {et~al.}(2017)\citenamefont
  {Cardarelli}, \citenamefont {Greschner},\ and\ \citenamefont
  {Santos}}]{Hidden2017}%
  \BibitemOpen
  \bibfield  {author} {\bibinfo {author} {\bibfnamefont {L.}~\bibnamefont
  {Cardarelli}}, \bibinfo {author} {\bibfnamefont {S.}~\bibnamefont
  {Greschner}}, \ and\ \bibinfo {author} {\bibfnamefont {L.}~\bibnamefont
  {Santos}},\ }\href {\doibase 10.1103/PhysRevLett.119.180402} {\bibfield
  {journal} {\bibinfo  {journal} {Phys. Rev. Lett.}\ }\textbf {\bibinfo
  {volume} {119}},\ \bibinfo {pages} {180402} (\bibinfo {year}
  {2017})}\BibitemShut {NoStop}%
\bibitem [{\citenamefont {Xu}\ \emph {et~al.}(2018)\citenamefont {Xu},
  \citenamefont {Gu},\ and\ \citenamefont {Mueller}}]{Hidden2018}%
  \BibitemOpen
  \bibfield  {author} {\bibinfo {author} {\bibfnamefont {J.}~\bibnamefont
  {Xu}}, \bibinfo {author} {\bibfnamefont {Q.}~\bibnamefont {Gu}}, \ and\
  \bibinfo {author} {\bibfnamefont {E.~J.}\ \bibnamefont {Mueller}},\ }\href
  {\doibase 10.1103/PhysRevLett.120.085301} {\bibfield  {journal} {\bibinfo
  {journal} {Phys. Rev. Lett.}\ }\textbf {\bibinfo {volume} {120}},\ \bibinfo
  {pages} {085301} (\bibinfo {year} {2018})}\BibitemShut {NoStop}%
\end{thebibliography}

\begin{thebibliography}{7}%
\makeatletter
\providecommand \@ifxundefined [1]{%
 \@ifx{#1\undefined}
}%
\providecommand \@ifnum [1]{%
 \ifnum #1\expandafter \@firstoftwo
 \else \expandafter \@secondoftwo
 \fi
}%
\providecommand \@ifx [1]{%
 \ifx #1\expandafter \@firstoftwo
 \else \expandafter \@secondoftwo
 \fi
}%
\providecommand \natexlab [1]{#1}%
\providecommand \enquote  [1]{``#1''}%
\providecommand \bibnamefont  [1]{#1}%
\providecommand \bibfnamefont [1]{#1}%
\providecommand \citenamefont [1]{#1}%
\providecommand \href@noop [0]{\@secondoftwo}%
\providecommand \href [0]{\begingroup \@sanitize@url \@href}%
\providecommand \@href[1]{\@@startlink{#1}\@@href}%
\providecommand \@@href[1]{\endgroup#1\@@endlink}%
\providecommand \@sanitize@url [0]{\catcode `\\12\catcode `\$12\catcode
  `\&12\catcode `\#12\catcode `\^12\catcode `\_12\catcode `\%12\relax}%
\providecommand \@@startlink[1]{}%
\providecommand \@@endlink[0]{}%
\providecommand \url  [0]{\begingroup\@sanitize@url \@url }%
\providecommand \@url [1]{\endgroup\@href {#1}{\urlprefix }}%
\providecommand \urlprefix  [0]{URL }%
\providecommand \Eprint [0]{\href }%
\providecommand \doibase [0]{http://dx.doi.org/}%
\providecommand \selectlanguage [0]{\@gobble}%
\providecommand \bibinfo  [0]{\@secondoftwo}%
\providecommand \bibfield  [0]{\@secondoftwo}%
\providecommand \translation [1]{[#1]}%
\providecommand \BibitemOpen [0]{}%
\providecommand \bibitemStop [0]{}%
\providecommand \bibitemNoStop [0]{.\EOS\space}%
\providecommand \EOS [0]{\spacefactor3000\relax}%
\providecommand \BibitemShut  [1]{\csname bibitem#1\endcsname}%
\let\auto@bib@innerbib\@empty
\bibitem [{\citenamefont {Majumdar}\ and\ \citenamefont
  {Ghosh}(1969)}]{Majumdar1969}%
  \BibitemOpen
  \bibfield  {author} {\bibinfo {author} {\bibfnamefont {C.~K.}\ \bibnamefont
  {Majumdar}}\ and\ \bibinfo {author} {\bibfnamefont {D.~K.}\ \bibnamefont
  {Ghosh}},\ }\href {\doibase 10.1063/1.1664979} {\bibfield  {journal}
  {\bibinfo  {journal} {Journal of Mathematical Physics}\ }\textbf {\bibinfo
  {volume} {10}},\ \bibinfo {pages} {1399} (\bibinfo {year}
  {1969})}\BibitemShut {NoStop}%
\bibitem [{\citenamefont {Furukawa}\ \emph {et~al.}(2012)\citenamefont
  {Furukawa}, \citenamefont {Sato}, \citenamefont {Onoda},\ and\ \citenamefont
  {Furusaki}}]{Furukawa2012}%
  \BibitemOpen
  \bibfield  {author} {\bibinfo {author} {\bibfnamefont {S.}~\bibnamefont
  {Furukawa}}, \bibinfo {author} {\bibfnamefont {M.}~\bibnamefont {Sato}},
  \bibinfo {author} {\bibfnamefont {S.}~\bibnamefont {Onoda}}, \ and\ \bibinfo
  {author} {\bibfnamefont {A.}~\bibnamefont {Furusaki}},\ }\href {\doibase
  10.1103/PhysRevB.86.094417} {\bibfield  {journal} {\bibinfo  {journal} {Phys.
  Rev. B}\ }\textbf {\bibinfo {volume} {86}},\ \bibinfo {pages} {094417}
  (\bibinfo {year} {2012})}\BibitemShut {NoStop}%
\bibitem [{\citenamefont {Ueda}\ and\ \citenamefont
  {Onoda}(2014{\natexlab{a}})}]{Ueda2014Chiral}%
  \BibitemOpen
  \bibfield  {author} {\bibinfo {author} {\bibfnamefont {H.}~\bibnamefont
  {Ueda}}\ and\ \bibinfo {author} {\bibfnamefont {S.}~\bibnamefont {Onoda}},\
  }\href {\doibase 10.1103/PhysRevB.89.024407} {\bibfield  {journal} {\bibinfo
  {journal} {Phys. Rev. B}\ }\textbf {\bibinfo {volume} {89}},\ \bibinfo
  {pages} {024407} (\bibinfo {year} {2014}{\natexlab{a}})}\BibitemShut
  {NoStop}%
\bibitem [{\citenamefont {Ueda}\ and\ \citenamefont
  {Onoda}(2014{\natexlab{b}})}]{Ueda2014Chiral2}%
  \BibitemOpen
  \bibfield  {author} {\bibinfo {author} {\bibfnamefont {H.}~\bibnamefont
  {Ueda}}\ and\ \bibinfo {author} {\bibfnamefont {S.}~\bibnamefont {Onoda}},\
  }\href {\doibase 10.1103/PhysRevB.90.214425} {\bibfield  {journal} {\bibinfo
  {journal} {Phys. Rev. B}\ }\textbf {\bibinfo {volume} {90}},\ \bibinfo
  {pages} {214425} (\bibinfo {year} {2014}{\natexlab{b}})}\BibitemShut
  {NoStop}%
\bibitem [{\citenamefont {Calabrese}\ and\ \citenamefont
  {Cardy}(2004)}]{Cardy2004}%
  \BibitemOpen
  \bibfield  {author} {\bibinfo {author} {\bibfnamefont {P.}~\bibnamefont
  {Calabrese}}\ and\ \bibinfo {author} {\bibfnamefont {J.}~\bibnamefont
  {Cardy}},\ }\href {http://stacks.iop.org/1742-5468/2004/i=06/a=P06002}
  {\bibfield  {journal} {\bibinfo  {journal} {Journal of Statistical Mechanics:
  Theory and Experiment}\ }\textbf {\bibinfo {volume} {2004}},\ \bibinfo
  {pages} {P06002} (\bibinfo {year} {2004})}\BibitemShut {NoStop}%
\bibitem [{\citenamefont {Giamarchi}(2003)}]{Giamarchi2003}%
  \BibitemOpen
  \bibfield  {author} {\bibinfo {author} {\bibfnamefont {T.}~\bibnamefont
  {Giamarchi}},\ }\href {\doibase 10.1093/acprof:oso/9780198525004.001.0001}
  {\emph {\bibinfo {title} {Quantum Physics in One Dimension}}}\ (\bibinfo
  {publisher} {Oxford University Press},\ \bibinfo {year} {2003})\BibitemShut
  {NoStop}%
\bibitem [{\citenamefont {Vekua}\ \emph {et~al.}(2003)\citenamefont {Vekua},
  \citenamefont {Japaridze},\ and\ \citenamefont {Mikeska}}]{FMleg}%
  \BibitemOpen
  \bibfield  {author} {\bibinfo {author} {\bibfnamefont {T.}~\bibnamefont
  {Vekua}}, \bibinfo {author} {\bibfnamefont {G.~I.}\ \bibnamefont
  {Japaridze}}, \ and\ \bibinfo {author} {\bibfnamefont {H.-J.}\ \bibnamefont
  {Mikeska}},\ }\href {\doibase 10.1103/PhysRevB.67.064419} {\bibfield
  {journal} {\bibinfo  {journal} {Phys. Rev. B}\ }\textbf {\bibinfo {volume}
  {67}},\ \bibinfo {pages} {064419} (\bibinfo {year} {2003})}\BibitemShut
  {NoStop}%
\end{thebibliography}
%

\pagebreak
\widetext
\begin{center}
\textbf{\large Supplemental Materials for ``Exactly solvable points and symmetry protected topological phases of quantum spins on a zig-zag lattice"}
\end{center}
\begin{center}
\textbf{Haiyuan Zou, Erhai Zhao, Xi-Wen Guan, and W. Vincent Liu}
\end{center}

\setcounter{equation}{0}
\setcounter{figure}{0}
\setcounter{table}{0}
\setcounter{page}{1}
\makeatletter
\renewcommand{\thefigure}{S\arabic{figure}}
\renewcommand{\thetable}{S\arabic{table}}
\renewcommand{\theequation}{S\arabic{equation}}
\renewcommand{\bibnumfmt}[1]{[S#1]}
\renewcommand{\citenumfont}[1]{S#1}
\makeatother

\section{Exact solution}
We rewrite the Hamiltonian Eq (1) in the main text in terms of Pauli matrices:

\begin{equation}
H=\frac{1}{4}\sum_{i,j}J_{i,j}(\sigma^x_i\sigma^x_j+\sigma^y_i\sigma^y_j+\eta\sigma^z_i\sigma^z_j),
\end{equation}
and consider the case $J_{2i,2i+1}=J_1>0,J_{2i-1,2i}=J_1'<0$, and $J_{i,i+2}=J_2<0$.

Defining the local Hamiltonian
\begin{equation}
\tilde{H}_{i,j}=\frac{1}{2}J_{i,j}(1-\sigma^x_i\sigma^x_j-\sigma^y_i\sigma^y_j-\eta\sigma^z_i\sigma^z_j),
\end{equation}
and using it to rewrite the total Hamiltonian with $2N$ sites and periodic boundary condition as:
\begin{equation}
H=\frac{1}{4}N(J_1+J_1'+2J_2)-\frac{1}{2}\sum_{i,j}\tilde{H}_{i,j}.
\end{equation}

Following Majumdar and Ghosh's notation~\cite{Majumdar1969}, we define singlet product state 
$|\psi\rangle_s=[1,2][3,4]\dots[2N-1,2N]$, 
and even-parity prodect state $|\psi\rangle_e=\{2,3\}\{4,5\}\dots\{2N,1\}$,
in which 
$[i,j]$ represents a singlet $(|\uparrow\rangle_i|\downarrow\rangle_j-|\downarrow\rangle_i|\uparrow\rangle_j)/\sqrt{2}$ associated with site $i,j$ and $\{i,j\}$ is a even-parity state $(|\uparrow\rangle_i|\downarrow\rangle_j+|\downarrow\rangle_i|\uparrow\rangle_j)/\sqrt{2}$.

By straightforward calculation, we find
\begin{eqnarray}
\tilde{H}_{i,j}[i,j]/J_{i,j}&=&\frac{3+\eta}{2}[i,j],\\
\tilde{H}_{i,j}\{i,j\}/J_{i,j}&=&\frac{\eta-1}{2}\{i,j\},\\
\tilde{H}_{i,j}[k,i][j,n]/J_{i,j}&=&[i,j][n,k]-\frac{1-\eta}{2}\{k,i\}\{j,n\},\\
\tilde{H}_{i,j}\{k,i\}\{j,n\}/J_{i,j}&=&[i,j][n,k]-\frac{1-\eta}{2}[k,i][j,n],\\
\tilde{H}_{i,j}\{k,i\}[j,n]/J_{i,j}&=&-[i,j]\{n,k\}-\frac{1-\eta}{2}[k,i]\{j,n\},
\end{eqnarray}
and using the algebraic identity
\begin{equation}
[k,l][m,n]+[k,n][l,m]+[k,m][n,l]=0,
\end{equation}
we can obtain that
\begin{eqnarray}
\nonumber
\tilde{H}_{1,2}|\psi\rangle_s&=&J_1\frac{3+\eta}{2}|\psi\rangle_s,\\ \nonumber
\tilde{H}_{2,3}|\psi\rangle_s&=&J_1'([2,3][4,1]-\frac{1-\eta}{2}\{1,2\}\{3,4\})[5,6]\dots[2N-1,N],\\ \nonumber
\dots&&,\\ \nonumber
\tilde{H}_{1,3}|\psi\rangle_s&=&\tilde{H}_{2,4}|\psi\rangle_s \\ \nonumber
&=&J_2([1,2][3,4]-[2,3][4,1]+\frac{1-\eta}{2}\{1,2\}\{3,4\})[5,6]\dots[2N-1,N].
\end{eqnarray}
Thus,
\begin{equation}
H|\psi\rangle_s=-\frac{2+\eta}{4}NJ_1|\psi\rangle_s-\frac{J_1'-2J_2}{2}([2,3][4,1]-\frac{1-\eta}{2}\{1,2\}\{3,4\}[5,6]...[2N-1,2N]+...),
\end{equation}
which means that, at $J_1'=2J_2$, the singlet product state is the eigenstate of the total Hamiltonian with the average energy $E_s=E/(2N)=-\frac{2+\eta}{8}J_1$. Using the definition of the couplings in Eq.~(3) in the main text, for $\gamma=30^\circ$, one can obtain that at $\theta= 50.9^\circ$, the relation $J_1'=2J_2$ is satisfied.

Similarly, for the even-parity product state $|\psi\rangle_e$
\begin{eqnarray}
\nonumber
H|\psi\rangle_e&=&\frac{N}{4}[J_1'(2-\eta)+J_1+2J_2]|\psi\rangle_e\\
&-&\frac{J_1+2J_2}{2}([3,4][5,2]...[2N,1]+...)\\ \nonumber
&+&(J_2\frac{1+\eta}{2}+J_1\frac{1-\eta}{4})([2,3][4,5]...[2N,1]+...).
\end{eqnarray}
We can prove that at $J_1=-2J_2$ and $\eta=0$, the even-parity product state is an eigenstate of the Hamiltonian, with the average energy $E_e=E/(2N)=J_1'/4$. For $\gamma=30^\circ$, this corresponds to a particular point at $\theta=42.4^\circ$.

To further prove that these eigenstates are the ground state of the Hamiltonian, Rayleigh-Ritz inequality $E_{g.s}\equiv\langle\psi_{g.s}|H|\psi_{g.s}\rangle=\langle\psi_{g.s}|\sum_iH_i|\psi_{g.s}\rangle\ge\sum_iE_i$ is used. The ground state energy $E_{g.s}$ of the total system $H$ is not less than the summation of the ground state energy $E_i$ of each ingredient part $H_i$. Thus, once the eigenenergy $E_s$ or $E_e$ is the same with $E_i$, the eigenstate is also the ground state. 

The zigzag chain Hamiltonian with $2N$ sites can be decomposed into summation of $2N$ small triangle Hamiltonians $H=\sum_iH_i$, where the Hamiltonian for a single triangle labeled by $i$ is
\begin{equation}
 H_i=J_1h_{i,i+1}/8+J_1'h_{i+1,i+2}/8+J_2h_{i,i+2}/4,
 \end{equation} 
where $h_{ij}=\sigma^x_i\sigma^x_{j}+\sigma^y_i\sigma^y_{j}+\eta\sigma^z_i\sigma^z_{j}$.
 
It is easy to calculate that the first two lowest eigenvalues at the limit $J_1'=2J_2$ are
\begin{eqnarray}
E_1&=&-\frac{2+\eta}{8}J_1,\\
E_2&=&\frac{1}{4}(J_1-\eta J_1')-\frac{1}{4}\sqrt{8J_1'^2+[J_1(1-\eta)+J_1'\eta]^2},
\end{eqnarray}
where $E_1=E_s$.

For $E_1\le E_2$, we can conclude that $|\psi\rangle_s$ is the ground state of SD phase indeed. For the case $J_1'=2J_2$, or at $\gamma=30^\circ,\theta=50.9^\circ$, it gives
$\eta\geq\frac{|J_1'|}{J_1}-1=0.747$. Note that for $\eta<0.747$, the singlet product state may also be the ground state because the Rayleigh-Ritz inequality only gives the lower bound of the energy and strong quantum fluctuation in 1D will enlarge the actual singlet product state region. Our iTEBD calculation shows that the critical value of $\eta$ is around 0.695 [Fig.~6(a)]. 

Following the same procedure, for $J_1=-2J_2$ and $\eta=0$, the lowest two eigenvalues of a small triangle Hamiltonian are $E_1'=J_1'/4$ and $E_2'=-(J_1'+\sqrt{8J^2_1+J_1'^2})/8$. By solving $E_1'<E_2'$, we find the condition for the even-parity product state being the ground state of the whole system is $J_1'<-J_1$. At $\gamma=30^\circ,\theta=42.6^\circ$, this relation is satisfied.     

\section{solutions of open boundary condition cases}

To understand the short range entanglement feature of both SPT phases and their edge states, we calculate the wave functions of systems with open boundary condition. Starting from the exact solvable case for SD phase ($J_1'=2J_2<0, J_1>0$) and considering the large $N$ limit, we first demonstrate the results of a open chain with $2N+1$ sites and show the free spin on one end can be generated. Using the label convention of spins as starting from 0 to $2N$, and setting the coupling on the first bond as $J_1'$, we define the wavefunction of a chain with a free spin on the left end singlet product state for the rest as 
\begin{equation}
\label{eq:wavf0}
|\tilde{\psi}\rangle_{s0}=\sigma_0[1,2][3,4]...[2N-1,2N]\hspace{1cm}(\sigma=\uparrow,\downarrow)
\end{equation}
and calculate the first two terms of the Hamiltonian operated on the first three sites. Taking spin $\uparrow$ as an example, using 

\begin{eqnarray}
H_{0,1}\uparrow_0[1,2]/J_1'&=&-2\downarrow_0\uparrow_1\uparrow_2+\eta\uparrow_0\{1,2\},\\
H_{0,1}\uparrow_0\{1,2\}/J_1'&=&2\downarrow_0\uparrow_1\uparrow_2+\eta\uparrow_0[1,2],\\
H_{0,1}\downarrow_0\uparrow_1\uparrow_2/J_1'&=&\uparrow_0\{1,2\}-\uparrow_0[1,2]-\eta\downarrow_0\uparrow_1\uparrow_2,\\
H_{0,2}\uparrow_0[1,2]/J_2&=&2\downarrow_0\uparrow_1\uparrow_2-\eta\uparrow_0\{1,2\},\\
H_{0,2}\uparrow_0\{1,2\}/J_2&=&2\downarrow_0\uparrow_1\uparrow_2-\eta\uparrow_0[1,2],\\
H_{0,2}\downarrow_0\uparrow_1\uparrow_2/J_2&=&\uparrow_0\{1,2\}+\uparrow_0[1,2]-\eta\downarrow_0\uparrow_1\uparrow_2,
\end{eqnarray}
 and Eq. (S4-S8), we can express the ground state $|\tilde{\psi}_s\rangle$ and energy as $H|\tilde{\psi}\rangle_s=\tilde{E}_s|\tilde{\psi}\rangle_s$, where the total ground state energy is
 \begin{equation}
 \tilde{E}_s=-\frac{1}{4}\Big[J_1(2+\eta)N+\frac{J_2^2(2+\eta^2)}{NJ_1(2+\eta)}\Big]
 \end{equation}
 and one of the wave function is
 \begin{equation}
 \label{eq:wavf}
 |\tilde{\psi}\rangle_s=\Big\{\uparrow_0[1,2]-\frac{J_2\eta}{NJ_1(2+\eta)}\uparrow_0\{1,2\}+\frac{2J_2}{NJ_1(2+\eta)}\downarrow_0\uparrow_1\uparrow_2\Big\}[3,4]...[2N-1,2N]+O(\frac{1}{N^2}).
 \end{equation}

The state with all the spin flipped are degenerate with Eq.~\ref{eq:wavf}. In the large $N$ limit, it is obvious that only the $|\tilde{\psi}\rangle_{s0}$ (Eq.~\ref{eq:wavf0}) part is dominant. 

Considering the same chain with the coupling on the first bond as $J_1'$, following the same procedure to the exact solvable case for ED phase ($J_1=-2J_2>0, J_1'<0$, and $\eta=0$), one can easily get the corresponding total ground state energy as
\begin{equation}
 \tilde{E}_e=\frac{1}{4}\Big(2J_1'N+\frac{J_2^2}{NJ_1'}\Big)
 \end{equation}
 and the wave function is
 \begin{eqnarray}
 \label{eq:wavf2}
 \nonumber
 |\tilde{\psi}\rangle_e&=&\{0,1\}\{2,3\}...\{2N-4,2N-3\}\Big\{\{2N-2,2N-1\}\uparrow_{2N}\\
 &-&\frac{3J_2^2}{2N^2J_1'^2}[2N-2,2N-1]\uparrow_{2N}-\frac{J_2}{NJ_1'}\uparrow_{2N-2}\uparrow_{2N-1}\downarrow_{2N}\Big\}+O(\frac{1}{N^2}).
 \end{eqnarray}
 
 Again, the state with all spin flipped are degenerate with Eq.~\ref{eq:wavf2} and at the large $N$ limit, the state
 \begin{equation}
 |\tilde{\psi}\rangle_{e0}=\{0,1\}\{2,3\}...\{2N-2,2N-1\}\sigma_{2N}\hspace{1cm}(\sigma=\uparrow,\downarrow)
 \end{equation}
  with one free spin on the right end and even-parity product state for the rest are the dominant part. 
  The above explanation can be generated to all the other cases of different open chain structure. For chain with odd sites number and coupling on the first bond as $J_1$, the SD phase can be describe by the state with a free spin on the left end and singlet product state for the rest, while the corresponding ED state is a free spin on the right end with even-parity product state for the rest. For a large chain with even sites, if the first bond coupling is $J_1'$, the SD state have free spins on both end, while ED state forms perfect even-parity product state without edge modes. Similarly, if the first bond coupling is $J_1$, the ED state have free spins at the two ends instead. 

  At the exact solvable limit for both SD and ED cases, the entanglement vanishes. General cases of SD and ED phases have finite entanglement but adiabatically connect to the exact solvable cases and will become singular at the transition point, which is shown clearly from Fig.~3(b) in the main text.    

Defining $\mathcal{L}(\mathcal{R})=0,1$ as the number of free spins on the left (right) end, and $\bar{\mathcal{L}}(\bar{\mathcal{R}})$ is the opposite cases for $\mathcal{L}(\mathcal{R})$, the SPT transition from SD to ED phase can be represented as:
  \begin{equation}
  \mathcal{L}[[...]]\mathcal{R}\rightarrow\bar{\mathcal{L}}\{\{...\}\}\bar{\mathcal{R}}
  \end{equation}
  where $[[...]]$ ($\{\{...\}\}$) stands for the singlet (even-parity) product state in the middle. 

\section{infinite time-evolving block decimation (iTEBD) calculation}
\subsection{Chiral phase}

At $\theta=0^\circ$, the dipolar molecules have ferromagnetic nearest neighbor couplings ($J_1=J_1'<0$) and antiferromagnetic next nearest neighbor couplings ($J_2>0$) with $J_1=-1.25J_2$, which supports a large gapless vector chiral phase region in between two dimer phases on the easy-plane exchange anisotropy parameter line~\cite{Furukawa2012}. The chiral phase is characterized by the order parameter
\begin{equation}
\langle \hat{\kappa}^z\rangle = \frac{1}{N} \sum_i\langle (\hat{\vec{S}}_i \times \hat{ \vec{S} }_{i+1})^z \rangle.
\end{equation}

\begin{figure}[h]
\centering
\includegraphics[width=0.45\textwidth]{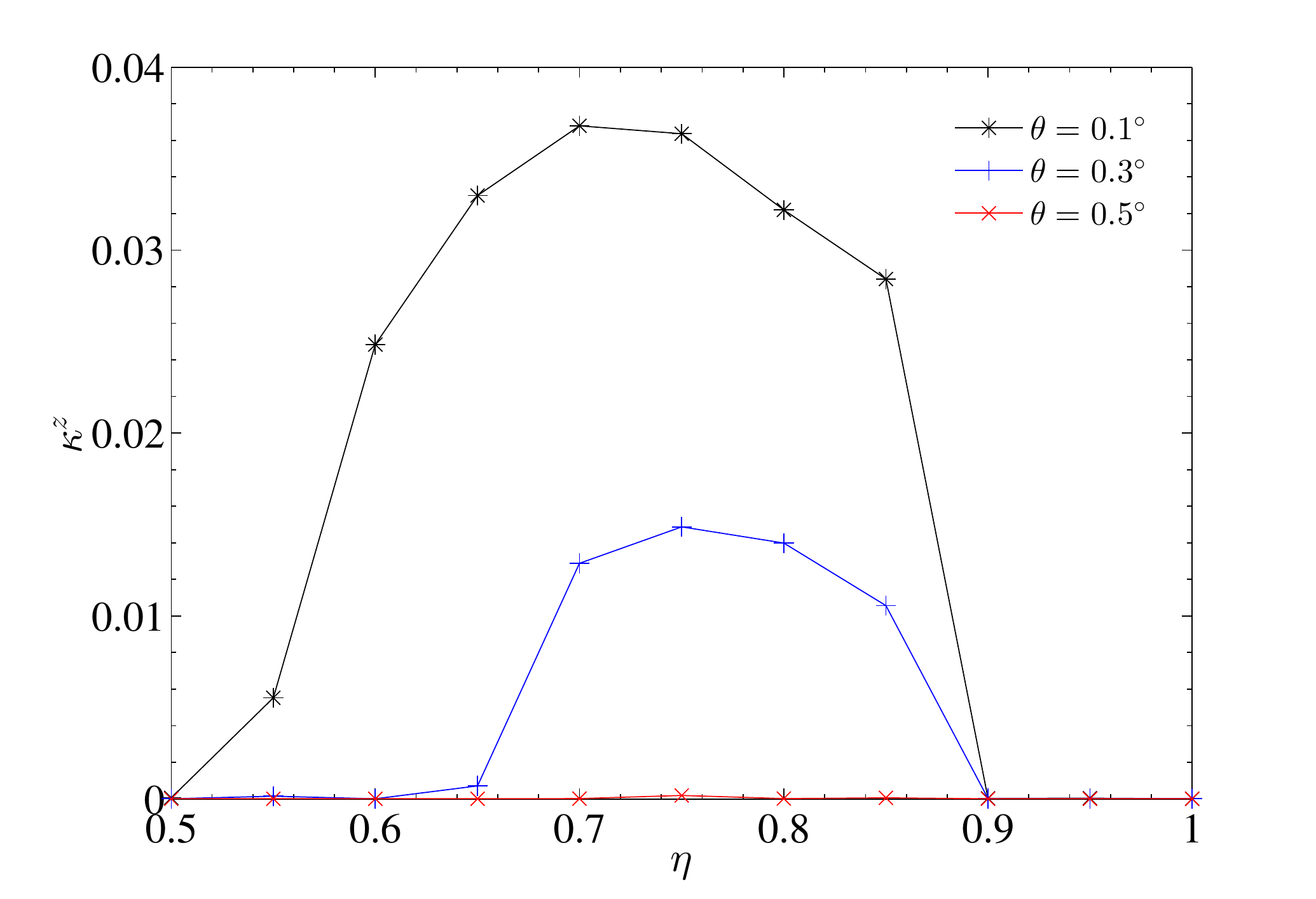}
\caption{Chiral order parameters as functions of $\eta$ at $\theta=0.1^\circ,0.3^\circ,$ and $0.5^\circ$, with $\chi=100$ are shown. At this tiny $\theta$ region, the chiral phase is suppressed with slightly increased $\theta$.}
\label{fig:chiral}
\end{figure}

 By introducing a tiny nearest bond alternation, the gap can be opened and the vector chiral order parameter is suppressed, which forms more SPT phases~\cite{Ueda2014Chiral,Ueda2014Chiral2}. This nearest bond alternation can be induced by slightly increasing $\theta$. Figure.~\ref{fig:chiral} shows that $\langle\kappa^z\rangle$ is negligible when $\theta$ increases only up to $0.5^\circ$.

\subsection{Other physical quantities on $\eta=1$ Heisenberg limit}

\begin{figure}
\centering
\includegraphics[width=0.45\textwidth]{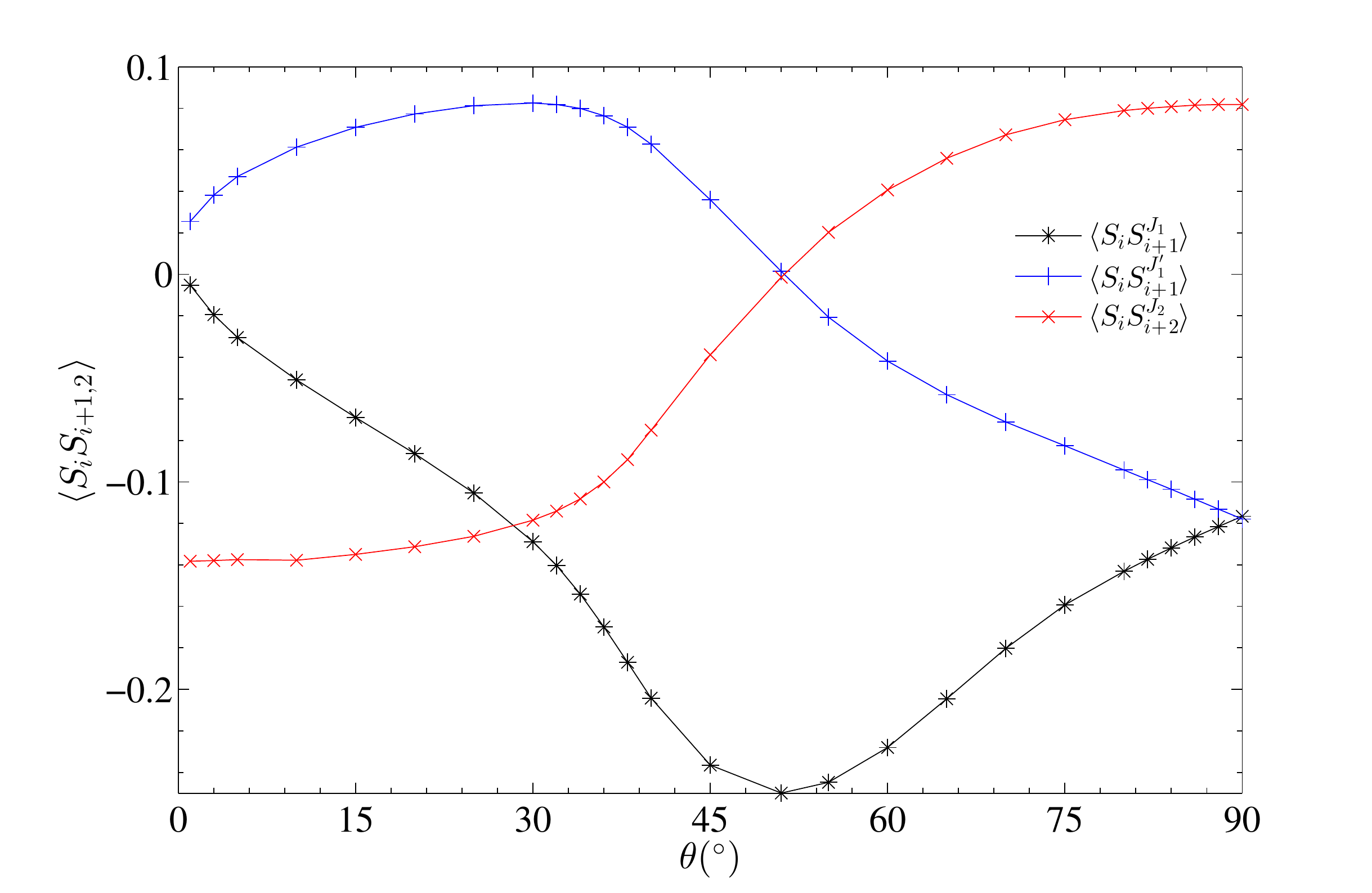}
\includegraphics[width=0.45\textwidth]{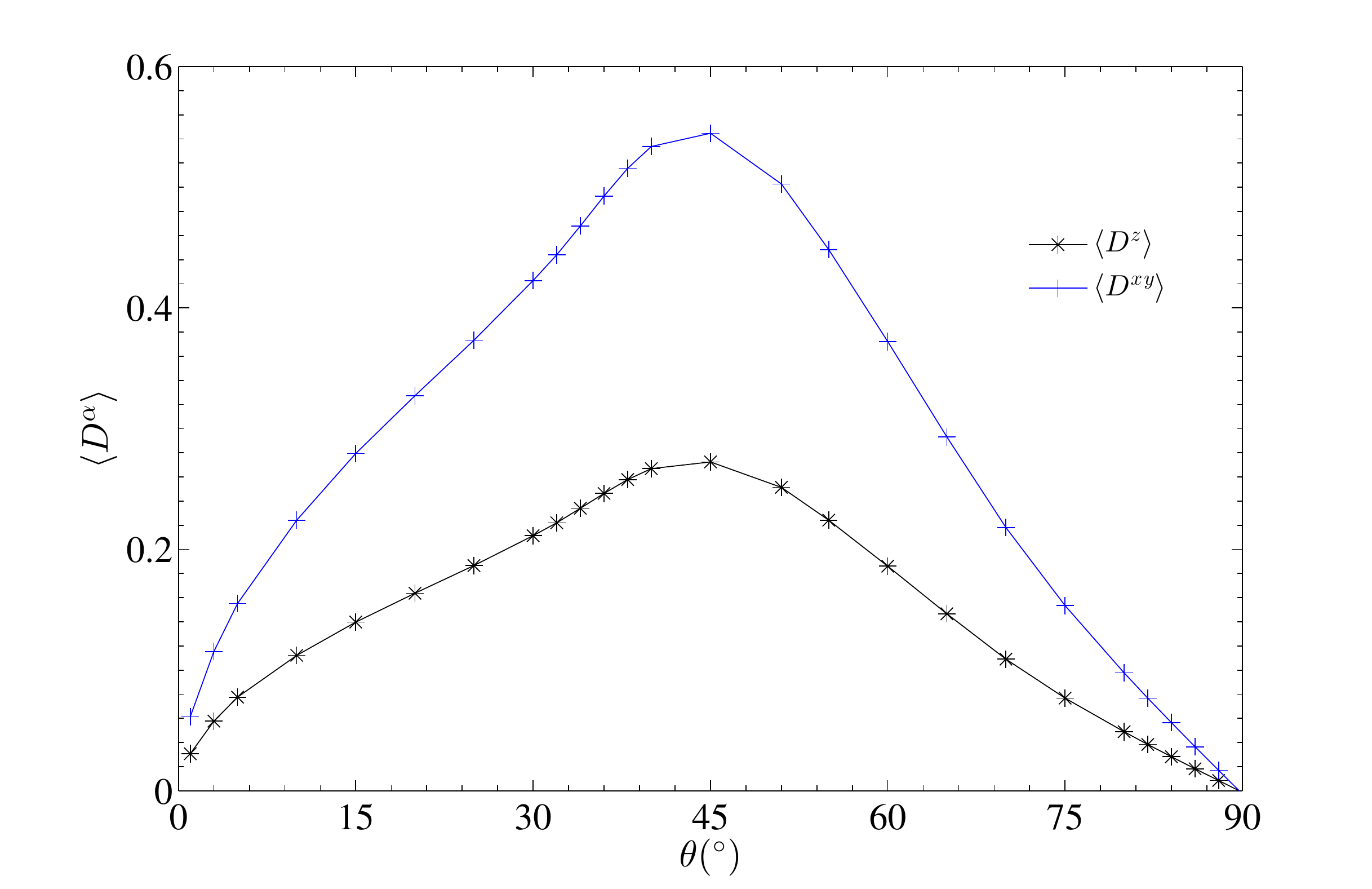}
\caption{Correlations on $J_1,J_1'$, and $J_2$ bond and dimer order parameters $D^z,D^{xy}$, for $\eta=1$ at varied $\theta$, with $\chi=300$ are shown. At the exact solvable point $\theta\sim 50.9^\circ$, only $\langle S_iS^{J_1}_{i+1}\rangle$ on $J_1$ is non-zero, suggest a pure singlet state on $J_1$ bond. }
\label{fig:cde}
\end{figure}

At the Heisenberg limit, any in-plane case are in a singlet dimer phase. This conclusion is farther checked by more physical quantities calculated by iTEBD. Fig.~\ref{fig:cde} shows that the bond correlations and the dimer order parameters are all smooth for continuous tuning of in-plane angle $\theta$, where the dimer order parameter is defined as,
\begin{equation}
\langle\hat{D}^\alpha\rangle=\frac{1}{N}\sum_i(-1)^{i-1}\langle\hat{S}^\alpha_{i}\hat{S}^\alpha_{i+1}\rangle.
\end{equation}
and $D^{xy}=D^x+D^y$ is the total in-plane dimer order parameter.

\subsection{SD to ED phase transition}
\begin{figure}
 \centering
 \includegraphics[width=0.45\textwidth]{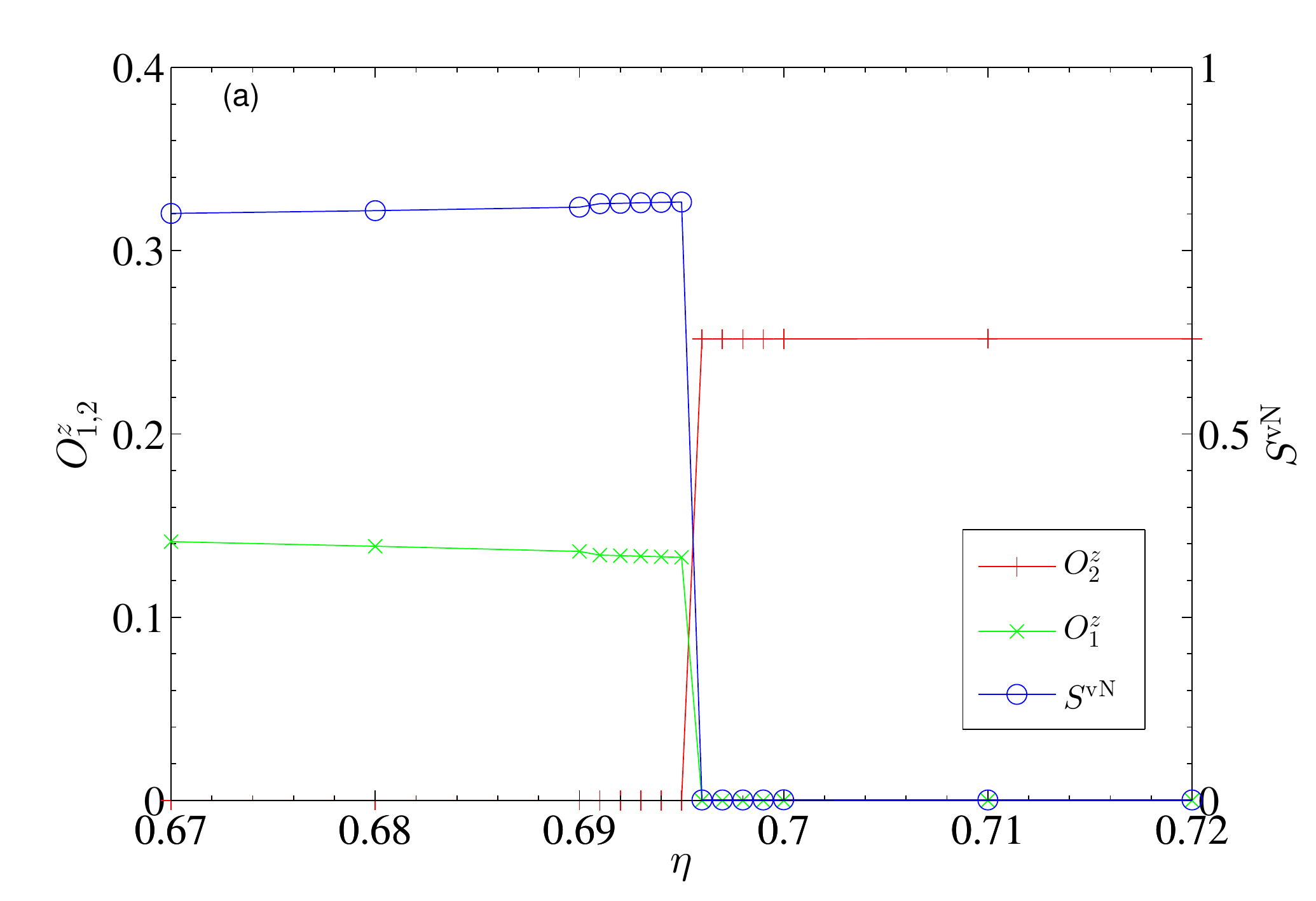}
 \includegraphics[width=0.45\textwidth]{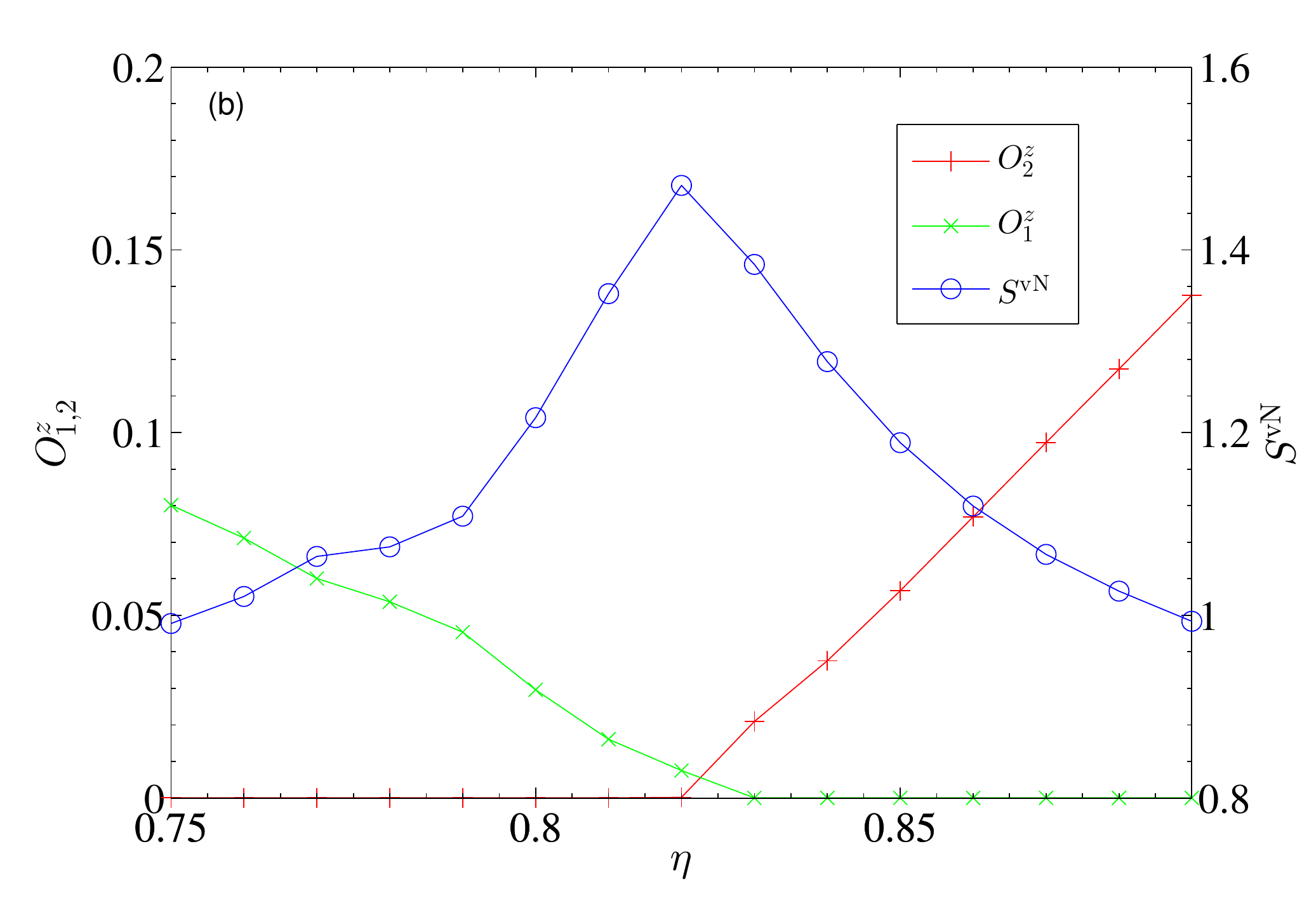}
 \caption{Entanglement Entropy ($S^{\rm vN}$) and singlet/even-parity string order parameter $O^z_{2/1}$ for (a) $\theta\sim 51^\circ$, (b) $\theta=30^\circ$, with $\chi=100$. The shapes of $O^z_{2/1}$ change at different $\theta$ suggests a continuously
varied critical exponents on the transition line from the SD phase to the ED phase.}
 \label{fig:difftheta}
 \end{figure}

Different from the previous studies~\cite{Furukawa2012,Ueda2014Chiral2} where a gapless vector chiral phase in between the SD and the ED phase, there is a direct phase transition between SD and ED phase in our system. We find that the physical properties on different points on the transition line from the SD to the ED phase are not unique. iTEBD calculation shows that at different $\theta$, the shapes of the string order parameters $O^z_{1/2}$ are continuously varied on the transition line [Fig.~\ref{fig:difftheta} and Fig.~3(b) in the main text], which suggest a phase transition with varied critical behaviors, similar with Gaussian type phase transition.

 \begin{figure}
 \centering
 \includegraphics[width=0.45\textwidth]{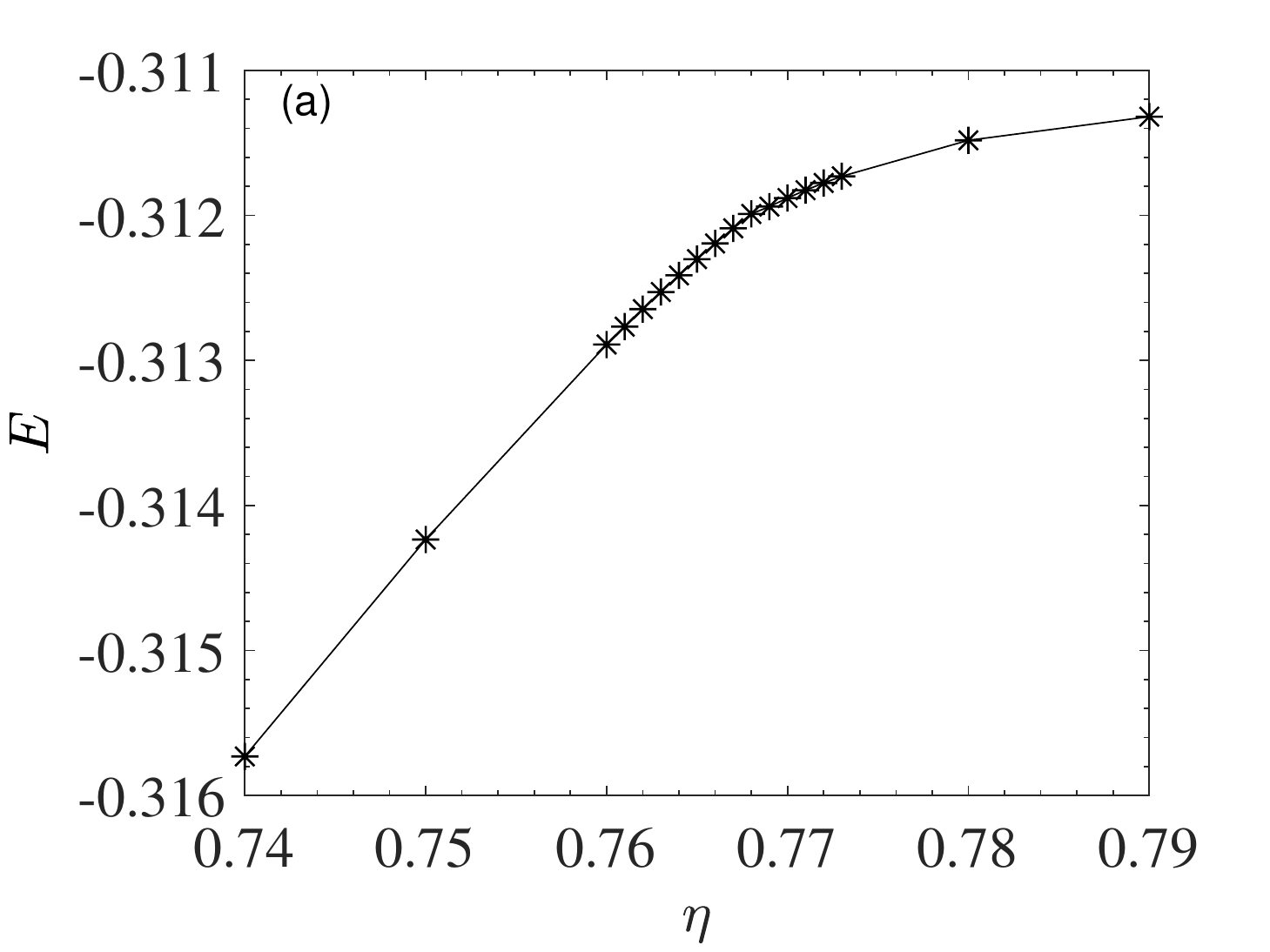}
 \includegraphics[width=0.45\textwidth]{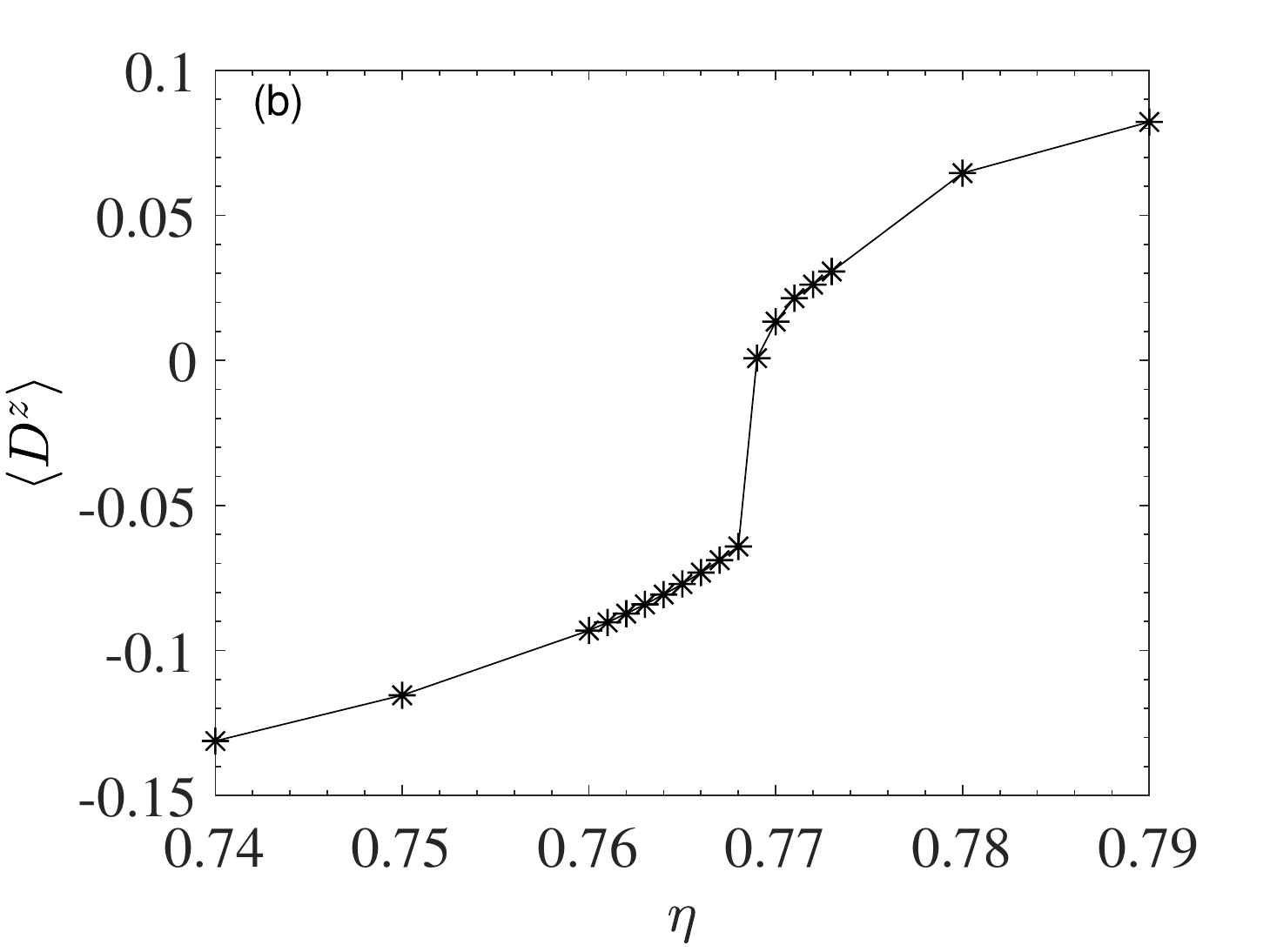}
 \includegraphics[width=0.45\textwidth]{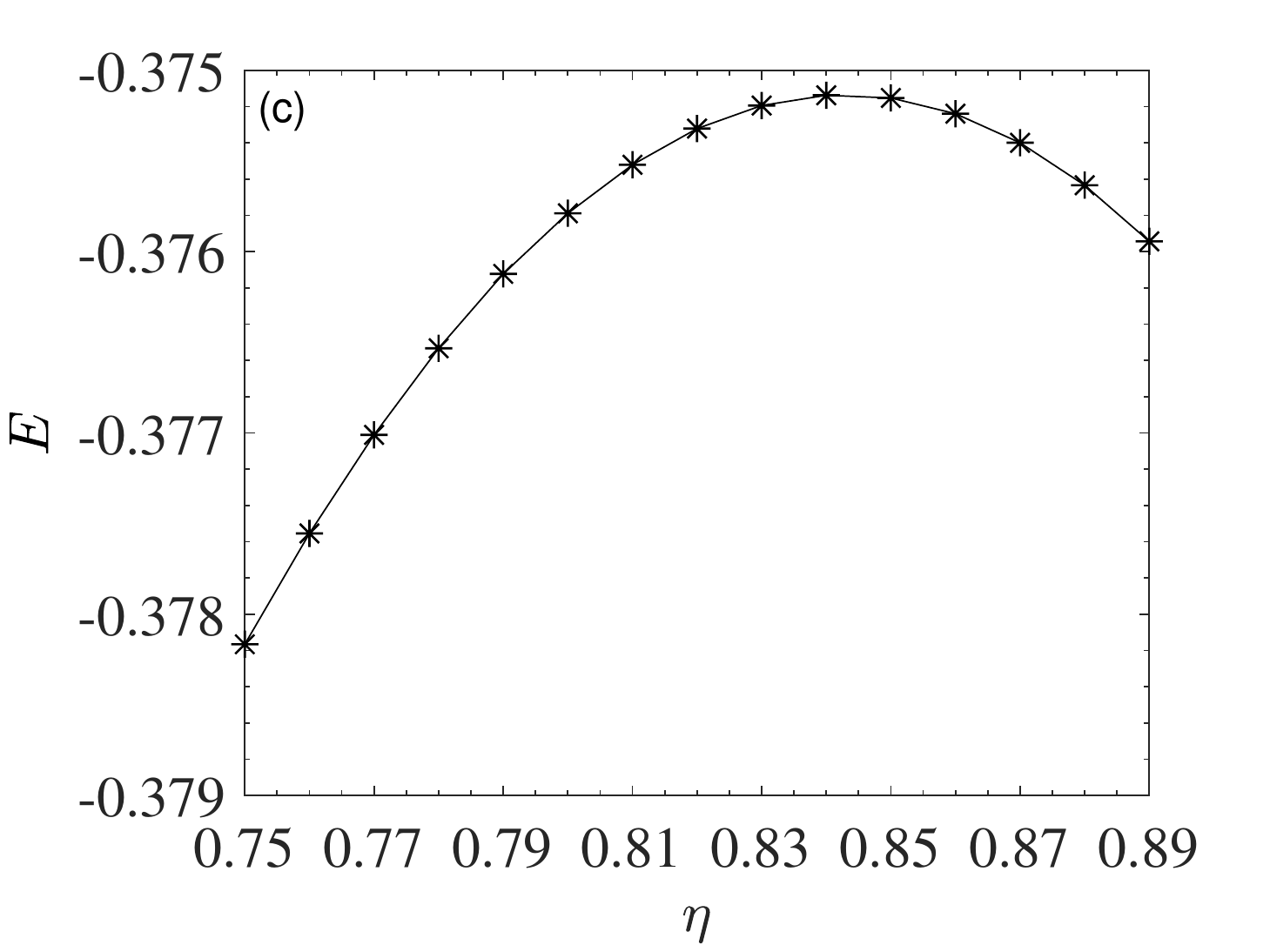}
 \includegraphics[width=0.45\textwidth]{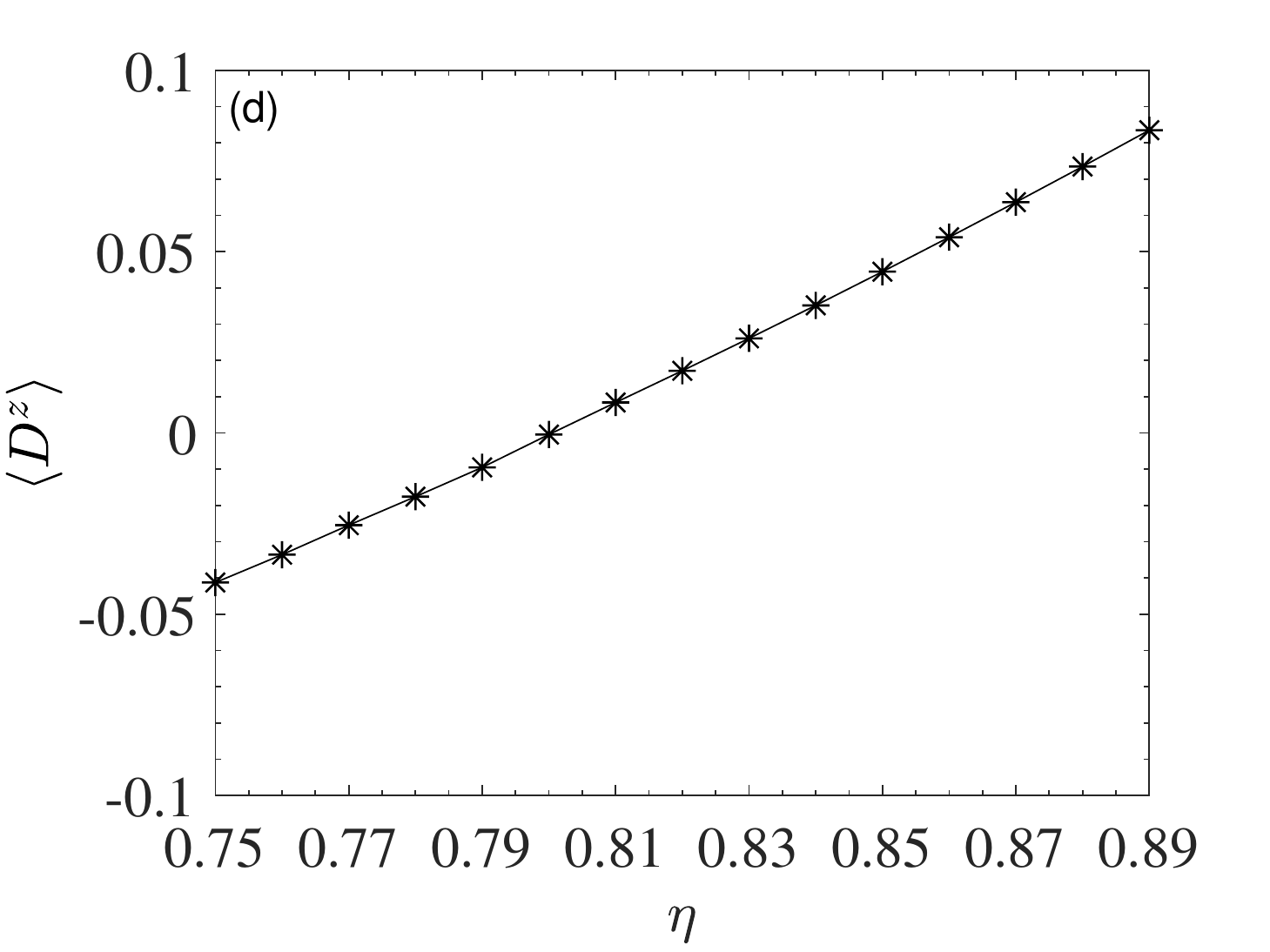}
 \caption{Results as functions of $\eta$ for $\theta=40^\circ$: (a) ground state energy $E$, (b) dimer order parameter $\langle D^z\rangle$ and for $\theta=30^\circ$: (c) ground state energy $E$, (d) dimer order parameter $\langle D^z\rangle$.}
 \label{fig:theta40}
 \end{figure}
At large $\theta$, for example, at the exact solvable point $\theta\sim50.9^\circ$, it is a strong first-order transition [Fig~3(a)]. At $\theta=40^\circ$, result from Fig.~3(b) in the main text suggests a first-order phase transition, but weaker than the transition at the exact solvable case. This is consistent with the results from ground state energy and dimer order parameter [Fig.~\ref{fig:theta40}(a,b)], the derivative of the energy or dimer order parameters are discontinuous at the phase transition point. As $\theta$ decreases, the transition becomes continuous, e.g. Fig.~\ref{fig:theta40}(c,d) shows that the energy and $D^z$ are smooth as $\eta$ is varied at $\theta=30^\circ$. 

\begin{figure}
\centering
\includegraphics[width=0.5\textwidth]{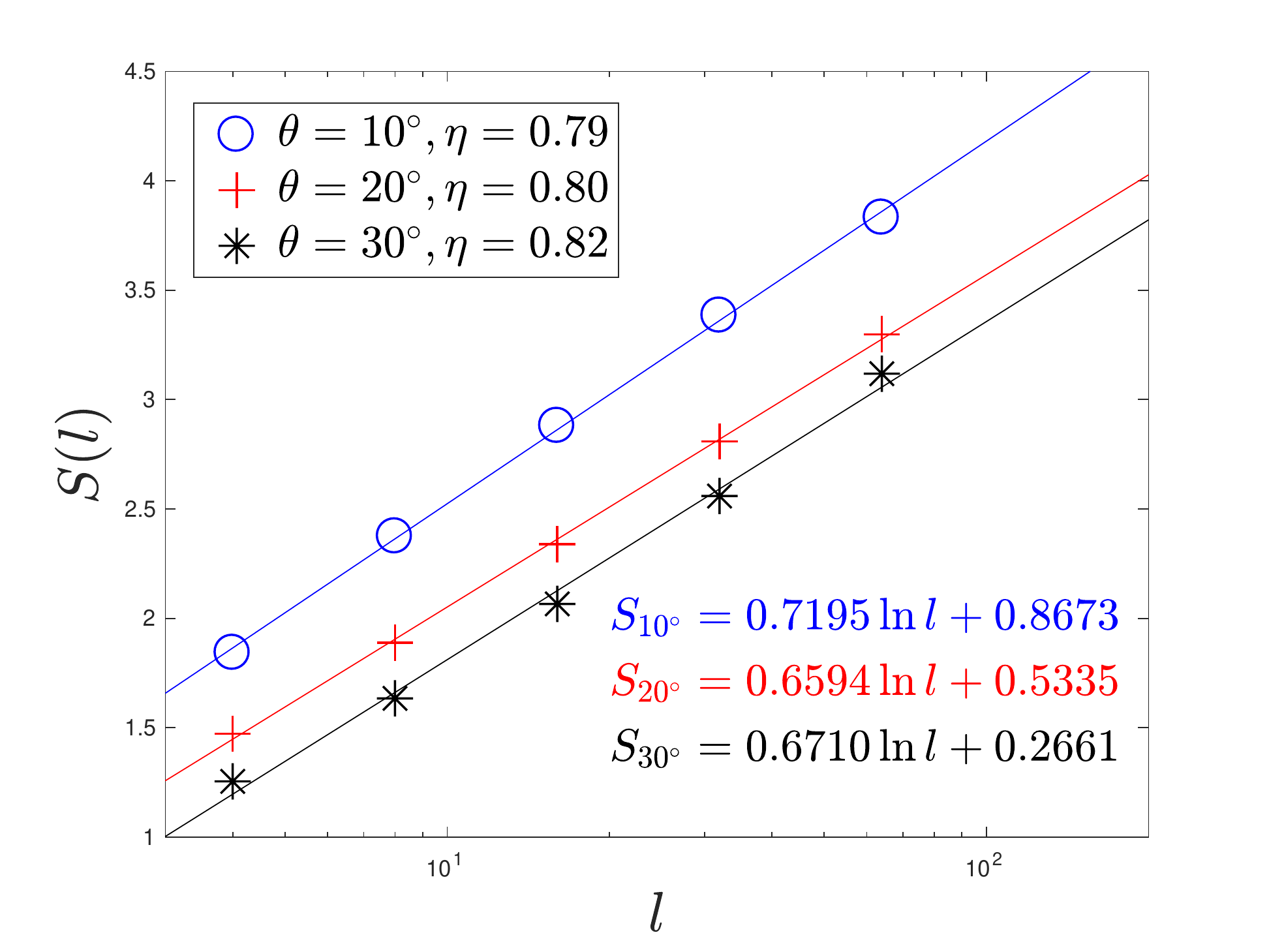}
\caption{The entanglement entropy $S(l)$ at the critical points as functions of interval distance $l$ in the semi-log scale for $\theta=10^\circ,20^\circ,30^\circ$. slope from linear fits for three cases give the central charge $c=2.16,1.98,2.01$ respectively.}
\label{fig:centralc}
\end{figure}

 To understand the critical behavior of this continuous phase transition and check if it is a Gaussian type phase transition, we calculate the central charge $c$ at the phase transition point at $\theta=10^\circ,20^\circ$, and $30^\circ$ by using the relation between the entanglement entropy and the site interval~\cite{Cardy2004}:
 \begin{equation}
 S(l)=\frac{c}{3}\ln l +\rm{const}
 \end{equation}
 where $S(l)$ is the entanglement entropy for a finite interval with length $l$. $S(l)$ can be calculated by $S(l)=\rm{Tr}\rho_l\ln\rho_l$, where $\rho_l$ is the reduced density matrix of the subsystem consisting $l$ sites and can be represented as $\rho_l=\rm{Tr}_m|\Psi\rangle\langle\Psi|$ ($|\Psi\rangle$ is the ground state), where the remainder subsystem $m$ is the traced. Results are shown in Fig.~\ref{fig:centralc}, linear fits give that $c\sim 2$ for three different cases, different from the Gaussian type phase transition where $c=1$. This large central charge suggests a much stronger interacting critical behavior for the SD to ED phase transition in general, different from the cases shown in Ref.~\cite{Ueda2014Chiral,Ueda2014Chiral2}, where $c=1$. 
 
 In our iTEBD calculation for the phase transition between SD and ED phase, we fix the bond dimension as $\chi=100$. Compare results from $\theta\sim50.9^\circ$ and $\theta=30^\circ$, the numerical error for the former is much smaller than the later case, which is shown in Fig.~\ref{fig:difftheta}. To get more precise results for smaller $\theta$ region, larger bond dimension is needed. We leave this quantitatively analysis for future study.
 
 \section{TLL to SD transition}
 For $\theta$ close to $90^\circ$, $J_1,J_1'\ll |J_2|$, model Eq.~(1) in the main text can be treated as two weakly coupled ferromagnetic XXZ chains, each being a Tomonaga-Luttinger liquid, using
abelian bosonization~\cite{Giamarchi2003}. From the bosonic fields $\phi_l(x)$ and their conjugates $\theta_l(x)$, where $[\theta_l(x), \phi_l(x')]=-i\pi\mathrm{\Theta}(x-x')$ and $l=1,2$ is the chain index, one constructs fields $\phi_\pm=(\phi_1\pm\phi_2)/\sqrt{2}$ and similarly $\theta_\pm$. Then the low-energy effective Hamiltonian density takes the form
$\mathcal{H}=\mathcal{H}_++\mathcal{H}_-+\mathcal{H}_{int}$, with
\begin{eqnarray}
\mathcal{H}_+&=&u_+K_+(\partial_x\theta_+)^2+\frac{u_+}{K_+}(\partial_x\phi_+)^2+g_1\cos(\sqrt{8}\phi_+),\nonumber \\
\mathcal{H}_-&=&u_-K_-(\partial_x\theta_-)^2+\frac{u_-}{K_-}(\partial_x\phi_-)^2+g_1\cos(\sqrt{8}\phi_-)\nonumber\\
&+&g_2\cos(\sqrt{2}\theta_-),\nonumber \\
\mathcal{H}_{int}&=&g_3\cos(\sqrt{2}\theta_-)\cos(\sqrt{8}\phi_+),
\label{eq:effect}
\end{eqnarray}
where the coupling constants $g_1\sim (J_1-J_1')\eta/\pi$, $g_2\sim J_1+J_1'$, $g_3\sim\ ({J_1-J_1'})/{2}$, and
$u_\pm=u \beta_\pm$, $K_\pm=K/\beta_\pm$, with $\beta_\pm= [1\pm {K(J_1+J_1')\eta}/(\pi u)]^{1/2}$.
The Luttinger parameter $K$ and velocity $u$ are given by
$K={\pi}/{(2\arccos\eta)}$, $u=|J_2| \sin(\pi/2K){K}/({2K-1})$.
From the renormalization group perspective, as $2-1/(2K_-)$ is always positive, the term $\cos(\sqrt{2}\theta_-)$ in $\mathcal{H}_-$ is relevant,
so the antisymmetric sector is gapped and we have condensation
$\Delta=\langle\cos(\sqrt{2}\theta_-)\rangle$ which depends on $\eta$. Then we can replace $\cos(\sqrt{2}\theta_-)$ with $\Delta$ in $\mathcal{H}_{int}$ and combine it with the $\cos(\sqrt{8}\phi_+)$ term
in $\mathcal{H}_+$, with $g_1\rightarrow g_1+g_3\Delta$. This leads to a sine-Gordon Hamiltonian $\mathcal{H}'=\mathcal{H}_++\mathcal{H}_{int}$ for the symmetric sector which can be analyzed following the standard procedure \cite{FMleg}. The term $\cos(\sqrt{8}\phi_+)$ becomes relevant when $2(K_+-1)<|g_1+g_3\Delta|$, and drives the TLL into the gapped SD phase via a Kosterlitz-Thouless transition. For $\eta\sim 0$, $J_1-J_1'\sim\eta$ while for $\eta\sim 1$, $J_1-J_1'\sim 1/(2\eta+\pi\Delta)$. Both suggest an arc-shaped phase boundary between the SD and TLL phase on the $\eta-\theta$ plane, consist with the iTEBD results.

\end{document}